\documentclass{emulateapj}

\shorttitle{AGN IN GROUPS AND CLUSTERS} 
\shortauthors{ARNOLD ET AL}
\slugcomment{Accepted by The Astrophysical Journal}

\newcommand{\chisqnu}{$\chi^{2}_{\nu}$}
\newcommand{\OIII}{$[\mathrm{O III}]\lambda5007/\mathrm{H}\beta$}
\newcommand{\NII}{$[\mathrm{N II}]\lambda6584/\mathrm{H}\alpha$}
\newcommand{\kms}{$\textrm{km s}^{-1}$}
\newcommand{\ergs}{$\textrm{erg s}^{-1}$}
\newcommand{\xmm}{{\it XMM-Newton}}

\begin{document}

\title{Active Galactic Nuclei in Groups and Clusters 
of Galaxies: Detection and Host Morphology} 

\author{Timothy J. Arnold\altaffilmark{1}, Paul Martini}

\affil{Department of Astronomy and Center for Cosmology and Astroparticle Physics, 
The Ohio State University, 
140 West 18th Avenue, 
Columbus, OH 43210;
arnold,martini@astronomy.ohio-state.edu}

\altaffiltext{1}{Current Address: 
Steward Observatory, University of Arizona,
933 N. Cherry Ave, 
Tucson, AZ 85721-0065; tjarnold@as.arizona.edu}

\author{John S. Mulchaey, Angela Berti}

\affil{Carnegie Observatories, 
813 Santa Barbara Street, 
Pasadena, CA 91101-1292;
mulchaey@ociw.edu}

\author{Tesla E. Jeltema}

\affil{UCO/Lick Observatories, 
1156 High Street,
Santa Cruz, CA 95064;
tesla@ucolick.org}

\begin{abstract}
The incidence and properties of Active Galactic Nuclei (AGN) in the field, 
groups, and clusters can provide new information about how these objects are 
triggered and fueled, similar to how these environments have been employed to 
study galaxy evolution. We have obtained new \xmm\ observations of seven 
X-ray selected groups and poor clusters with $0.02 < z < 0.06$ for comparison 
with previous samples that mostly included rich clusters and optically-selected 
groups. Our final sample has ten groups and six clusters in this low-redshift
range (split at a velocity dispersion of $\sigma = 500$ \kms).
We find that the X-ray selected AGN fraction increases 
from $f_A(L_X>10^{41}; M_R<M_R^*+1) = 0.047^{+0.023}_{-0.016}$ in clusters 
to $0.091^{+0.049}_{-0.034}$ for the groups (85\% significance), or a 
factor of two, for AGN above 
an 0.3-8keV X-ray luminosity of $10^{41}$\ergs\ hosted by galaxies more 
luminous than $M_R^*+1$. The trend is similar, although less significant, for 
a lower-luminosity host threshold of $M_R = -20$ mag. For many of the groups 
in the sample we have also identified AGN via standard emission-line 
diagnostics and find that these AGN are nearly disjoint from the X-ray 
selected AGN. Because there are substantial differences in the morphological 
mix of galaxies between groups and clusters, we have also measured the AGN 
fraction for early-type galaxies alone to determine if the differences are 
directly due to environment, or indirectly due to the change in the 
morphological mix. We find that the AGN fraction in early-type galaxies 
is also lower in clusters $f_{A,n>2.5}(L_X>10^{41}; M_R<M_R^*+1) 
= 0.048^{+0.028}_{-0.019}$ compared to $0.119^{+0.064}_{-0.044}$ for the 
groups (92\% significance), a result consistent with the hypothesis that the 
change in AGN fraction is directly connected to environment. 

\end{abstract}


\section{Introduction}

While it is clear that the Active Galactic Nuclei are powered by accretion onto 
supermassive black holes (SMBHs), and that this accretion requires both a 
source of fuel and a trigger to remove angular momentum from the fuel, the 
origin of the fuel and the trigger mechanism(s) remain poorly understood. 
For luminous QSOs, major mergers between gas-rich galaxies are considered the 
dominant mechanism and are the best, and perhaps only, candidate. Many studies 
have shown that a large fraction of these galaxies are morphologically 
disturbed, with close neighbors, tidal tails, multiple nuclei, or linked by 
luminous matter to other galaxies 
\citep{gehrens84,hutchings84,malkan84,smith86}. 
These results have motivated the hypothesis that lower-luminosity AGN in the
local Universe are, like QSOs, triggered and fueled by galaxy mergers or at 
least interactions, even though there is no evidence to support the claim that 
mergers are the trigger for low-luminosity AGN \citep{fuentes88,schmitt01}. 
If gas-rich mergers or interactions are the primary trigger for these 
lower-luminosity AGN, there should be higher AGN fractions in environments 
where galaxies have an abundant supply of gas and frequent interactions. 
While the cluster environment has very high number densities, the galaxies in 
the centers of rich clusters have less cold gas than those in less dense 
environments \citep[e.g.][]{giovanellihaynes85}. In addition, the high pairwise 
velocity dispersions of cluster members may be too large to allow the 
formation of bound pairs inside the virial radius \citep[e.g.][]{ghigna98}. In 
contrast, galaxies in the field have abundant supplies of cold gas, but the 
relatively low galaxy number density may counterbalance the abundance of fuel. 
Between these two extremes, the intermediate group environment could provide 
the ideal circumstances for the triggering and fueling of low-luminosity AGN 
in the nearby Universe, at least if mergers and interactions make a 
substantial contribution \citep[although see][]{martini04}. Galaxies in groups 
have sufficiently modest velocity dispersions, numerous neighbors, and 
available cold gas to trigger and fuel AGN. 

Many observations have shown that AGN are rarer in clusters when selected via 
emission-line diagnostics \citep{gisler78,dressler85}. Recent observations with 
the the Sloan Digital Sky Survey have shown this holds across a wide range in 
galaxy number density, specifically for the most luminous AGN 
\citep{miller03,kauffmann04,popesso06}. These results have readily been 
explained with the argument outlined above, namely that insufficient cold gas 
reservoirs and high velocity dispersions discourage the triggering and fueling 
of AGN in the cluster environment. However, these same observations have 
indicated that the differences between environments fade for less luminous AGN, 
and additionally samples of lower-luminosity AGN may vary substantially 
depending on the 
selection technique employed. The differences in member morphology and physical
environment between clusters and the field, 
as well as the relatively unbiased nature of X-ray observations, prompted 
\citet{martini06} to conduct a survey for X-ray AGN in nearby clusters. They 
found a factor of five higher AGN fraction than previous studies, where the 
AGN were identified to have X-ray luminosities $L_X \geq 10^{41}$ \ergs\ in 
host galaxies more luminous than $M_R = -20$ mag. This dramatic increase was 
attributed to the greater sensitivity of X-ray observations to lower-luminosity 
AGN relative to visible-wavelength emission-line diagnostics 
\citep{bpt81} because the contrast between X-ray emission 
from AGN and other physical processes (low-mass X-ray binaries, hot gas, and 
star formation) is higher than in the case of the emission-line diagnostics. 

To date only a small number of studies have attempted to extend work on 
X-ray selected AGN to the lower-velocity dispersion group environment. 
The first of these was a study of optically-selected groups at $z\sim0.06$ by 
\citet{shen07}. These authors only identified one AGN via X-ray selection 
out of 140 galaxies in eight groups, yet found five based on emission 
lines. More recently, \citet{sivakoff08} compared two, more massive groups and 
four clusters at similar redshifts and found a significantly higher 
X-ray selected AGN fraction in the groups compared to the clusters. One of 
the main goals of the present study is to dramatically improve on the 
small group sample in the \citet{sivakoff08} study, as well as to provide 
a larger sample of more massive groups than those in the \citet{shen07} 
study to better span the range of galaxy density from the field to 
clusters. While the groups in the \citet{shen07} sample are representative 
of those found in redshift surveys, they are otherwise a fairly heterogeneous 
sample. In contrast, the X-ray selection of this sample strongly suggests 
that these are all virialized systems. This local sample also provides a 
valuable benchmark for observations of the AGN fraction in groups and 
clusters at higher redshifts \citep[e.g.][]{jeltema07,silverman09,martini09}. 
For example, at $z=0.5-1$ \citet{georgakakis08b} find a similar fraction of 
X-ray selected AGN in optically-selected groups as \citet{shen07} in the local 
universe. 

Our other motivation is to investigate the role of galaxy morphology on the 
observed AGN fraction, as well as on the selection of AGN. Observations of 
many galaxies have shown that there is a strong correlation between the mass 
of the SMBH at the center of a given galaxy and the velocity dispersion or 
luminosity of the host galaxy's spheroid component 
\citep{ferrarese00,gebhardt00,marconi03}. These relationships imply the 
coevolution of SMBHs and spheroids, and many authors have suggested that 
AGN may actively impact the evolution of the host galaxy by quenching
star formation \citep{silk98,dimatteo05,hopkins05,springel05}. 
Since AGN are a consequence of matter accreting onto the central SMBH, and the 
relation between SMBH mass and bulge properties implies a connection in their 
evolution, there may be a relation between the incidence of AGN and galaxy 
morphology. Observations by \citet{ho97} with the nearby Palomar Seyfert 
Survey, and more recent work with SDSS \citep{kauffmann03}, indicate that this 
is the case. These studies find that Seyferts are preferentially found in 
early-type spirals with a significant bulge. 
One potential interpretation of this result is that because 
early-type galaxies of a given mass will have larger SMBHs compared to 
late-type galaxies of the same mass, if their SMBHs both accrete at the same 
fixed fraction of the Eddington rate, then the early-type galaxy is more 
likely to be detected in a luminosity-limited sample. 

This is important for comparing galaxy populations between groups, clusters, 
and the field because galaxy morphology is observed to be a strong function 
of local galaxy density \citep{dressler80}. Furthermore, many of the 
physical processes that are invoked to explain the relation between 
galaxy morphology and density may also have relevance for the available 
fuel supply for AGN. These include mergers and ram-pressure stripping via 
interactions with the hot ICM \citep{gunn72,quilis00}, evaporation of the 
cold ISM by the host ICM \citep{cowie77}, and starvation of new gas that would 
otherwise replenish the ISM \citep{larson80,balogh00}. There are thus 
correlations between morphology and the incidence of AGN, and between 
morphology and galaxy environment. It is therefore necessary to take 
morphology into account in order to determine if any variation in AGN fraction 
with environment is directly due to the environment itself, or indirectly 
due to the change in the mix of galaxy types with environment. 

To disentangle the effects of morphology and environment we present new 
observations of rich groups and poor clusters selected from the NORAS sample
of \citet{boehringer00}. These observations and the X-ray AGN classification 
procedure are described in \S\ref{sec:observations}. In addition, we use 
spectroscopic data from SDSS to classify AGN based on emission-line 
diagnostics and compare these AGN to those selected via X-ray emission. 
We then combine these new data with previous work \citep{martini06,sivakoff08} 
to study the incidence of AGN as a function of environment and morphology. 
The morphological classification is described in detail in 
\S\ref{sec:morphology}, and the AGN fractions are presented in 
\S\ref{sec:AGNFractions}. The results, including a statistical analysis, are 
described in \S\ref{sec:discussion}. The final section contains a summary 
of our main results. Throughout this paper we assume $H_0 = 70$ \kms\ 
Mpc$^{-1}$. 

\section{Observations and Sample Selection} \label{sec:observations}

\subsection{\xmm\ Observations} \label{sec:xmm} 

Previous studies of X-ray AGN and environment have concentrated on rich 
clusters \citep{martini06} and optically-selected groups \citep{shen07,
georgakakis08b}.  
X-ray detected groups represent the intermediate mass-scale between rich 
clusters and poor groups. To study the X-ray AGN population in X-ray detected 
groups, we observed a sample of seven low-redshift X-ray groups with the \xmm\ 
telescope. The groups were selected from the NORAS catalog, which provides a 
large, uniform sample of X-ray bright groups and clusters found in the ROSAT 
All-Sky Survey \citep{boehringer00}. Our groups 
were selected based on the following three criteria: 1) X-ray luminosities 
between $3 \times 10^{42}$ \ergs\ and $3 \times 10^{43}$ \ergs\ 
in the 0.1-2.4 keV band. This luminosity range was chosen to be comparable 
to the luminosity range of the intermediate redshift X-ray groups in the 
sample of \citet{mulchaey06}, potentially allowing a comparison of similar 
systems from redshift zero to $z \sim 0.5$; 2) Redshifts between 0.04 and 0.06. 
This redshift slice was chosen because it allows all of the groups to be 
studied out to approximately the virial radius in a single \xmm\ pointing; 
3) Spectroscopic coverage from the SDSS. SDSS spectroscopic coverage provides 
good membership information for each group. This last criterion also allows us 
to calculate an estimate of the AGN fraction based on standard emission-line 
diagnostics (see \S~\ref{sec:bptClassification} below) and to perform 
morphological fits of the surface brightness profiles of these objects from 
the SDSS imaging data (see \S~\ref{sec:morphology}). The above selection 
criteria result in a sample of 14 X-ray luminous groups and poor clusters. 
From these 14 groups, we selected seven that span the X-ray luminosity range 
of interest.

The seven X-ray groups were observed by \xmm\ between May 2007 and December 
2007. The integration times varied from 12 to 40 ksec. All observations were 
obtained in full-frame mode with the thin optical blocking filter. The data 
were processed with version 7.0 of the XMMSAS software. For EMOS data we used 
only patterns 0--12 and apply the \#XMMEA\_EM flag filtering, while for EPN
data we used patterns 0--4 and set the flag equal to zero. We eliminated 
periods of high background using the method described in \citet{jeltema06}.  
This procedure involves filtering the data in several energy bands. We begin 
by applying a cut on the high energy ($>$10 keV) count rate of 0.35 counts 
s$^{-1}$ for EMOS data and 1.0 counts s$^{-1}$ for EPN data. We then applied 
a $3\sigma$ clipping to the source-free count rate in three energy bands. For 
this process, we used time bins of length 100 seconds. Time bins with rates 
more than $3\sigma$ from the mean are then removed until the mean is stable. 
Background flaring was very severe for one of the groups (RXCJ1225.2+3213)
and resulted in no usable data. We therefore eliminated this group from
our sample. For the remaining six groups, the final exposure times were in 
the range $\sim$ 10 to 23 ksec for the EMOS detectors. We note that we only
include groups in our \xmm\ analysis for which we can reach a lower limit of
$L_X = 10^{41}$ \ergs\ at a radius of 13$'$.

To better constrain the AGN population in groups, we supplemented our sample 
with additional groups that have \xmm\ observations available from the archive. 
These groups were also selected from the NORAS catalog in the same X-ray 
luminosity range described above (criterion 1). We also required that 
these groups had spectroscopic coverage in the SDSS (criterion 3). However, we 
did not require these groups to be in the redshift range 0.04 to 0.06. To
insure that a significant fraction of the group members were within the \xmm\ 
field of view we only considered groups that had \xmm\ coverage out to a 
radius of at least 250 kpc. Using these revised criteria, we added another five 
groups to our sample, bringing the total number of NORAS-selected groups to 
eleven. The \xmm\ observations for these additional groups were reduced 
following the same method described above. The final exposure times for all 
five of these groups is at least 10 ksec in the EMOS detectors.

Membership was determined with the method described in \citet{mulchaey06} 
and largely based on the available SDSS spectroscopy in the fields on these 
groups and clusters. We start with all galaxies located within a projected 
distance of 1 Mpc from the center with a recessional velocity within $\pm 3000$ 
\kms\ of the mean velocity. We then calculate the velocity dispersion of the 
system using the biweight estimator \citep{beers90}. Objects with velocities 
greater than three times the velocity dispersion away from the mean are then 
removed from the sample and a new mean and dispersion are calculated. This 
process is continued until no further objects are removed. Note that although 
we have estimated the global properties from galaxies located within 1 Mpc of 
the center, our AGN analysis is restricted to the smaller radii probed by the 
\xmm\ data. 

\begin{deluxetable*}{llrrlrrlc}
\tablecolumns{9}
\tabletypesize{\scriptsize}
\tablecaption{Groups and Clusters} 
\tablehead{
\colhead{Group/Cluster Name} &
\colhead{Alternate Name} &
\colhead{RA} &
\colhead{Dec} &
\colhead{Redshift} &
\multicolumn{2}{c}{--- Members ---} &
\colhead{$\sigma_{v}$} &
\colhead{Data Sources} \\ 
\colhead{} &
\colhead{} &
\colhead{} &
\colhead{} &
\colhead{} &
\colhead{$N$} &
\colhead{$N^*$} &
\colhead{[km s$^{-1}$]} &
\colhead{} \\ 
 &  &  &  &  &  &  & \\                                                                                    
\colhead{(1)} &
\colhead{(2)} &
\colhead{(3)} &
\colhead{(4)} &
\colhead{(5)} &
\colhead{(6)} &
\colhead{(7)} &
\colhead{(8)} &
\colhead{(9)} 
}
\startdata
A85              &                  & 00:41:50.4  & -09:18:11  & 0.0554  & 109  & 53  & $993^{+85}_{-85}$   & 1,3 \\
A644             &                  & 08:17:25.6  & -07:30:45  & 0.0701  &  75  & 40  & $952^{+382}_{-382}$ & 2,3 \\ 
A3128            &                  & 03:30:43.8  & -52:31:30  & 0.0595  &  67  & 28  & $906^{+74}_{-74}$   & 2,3 \\
RXCJ0110.0+1358  &                  & 01:10:05.5  & +13:58:49  & 0.0581  &  30  & 15  & $745^{+74}_{-64} $  & 1,4 \\     
RXCJ0746.6+3100  & ZwCl0743.5+3110  & 07:46:37.3  & +31:00:49  & 0.0579  &  23  & 16  & $719^{+97}_{-59} $  & 1,4 \\   
RXCJ1022.0+3830  &                  & 10:22:04.7  & +38:30:43  & 0.0544  &  36  & 18  & $710^{+77}_{-54} $  & 1,4 \\  
 &  &  &  &  &  &  & \\
A3125            &                  & 03:27:17.9  & -53:29:37  & 0.0616  &  20  & 15  & $475^{+94}_{-94}$   & 2,3 \\
A89B             &                  & 00:42:54.6  & -09:13:50  & 0.0770  &  22  & 12  & $474^{+155}_{-155}$ & 1,3 \\
RXCJ0844.9+4258  &                  & 08:44:56.7  & +42:58:54  & 0.0550  &  13  &  9  & $343^{+75}_{-34} $  & 1,4 \\  
RXCJ1002.6+3241  & ZwCl0959.6+3257  & 10:02:38.6  & +32:41:58  & 0.0505  &  33  &  9  & $454^{+58}_{-39} $  & 1,4 \\   
RXCJ1122.2+6712  &                  & 11:22:14.5  & +67:12:46  & 0.0553  &  22  &  8  & $223^{+27}_{-23} $  & 1,4 \\   
RXCJ1204.4+0154  & MKW4             & 12:04:25.6  & +01:54:04  & 0.0203  &  12  &  7  & $495^{+59}_{-45} $  & 1,4 \\
RXCJ1223.1+1037  & NGC4325          & 12:23:06.5  & +10:37:26  & 0.0255  &   4  &  2  & $334^{+51}_{-32} $  & 1,4 \\   
RXCJ1324.1+1358  & NGC5129          & 13:24:11.9  & +13:58:45  & 0.0233  &   6  &  3  & $303^{+47}_{-29} $  & 1,4 \\  
RXCJ1440.6+0328  & MKW8             & 14:40:38.2  & +03:28:25  & 0.0269  &  15  &  9  & $449^{+41}_{-30} $  & 1,4 \\    
RXCJ1604.9+2355  & AWM4             & 16:04:57.0  & +23:55:14  & 0.0321  &   9  &  3  & $423^{+58}_{-36} $  & 1,4 
\enddata
\tablecomments{Properties of all groups and clusters employed in this study. 
Columns are: (1) Name in the original NORAS catalog or from \citet{sivakoff08}; 
(2) Alternate name from the literature, if any; (3 and 4) RA and DEC of the 
X-ray center; (5) redshift; (6 and 7) number of members more luminous than 
$M_R = -20$ and $M_R^*+1$, respectively; (8) velocity dispersion; (9) 
references for data. References are 1: SDSS DR6 images and spectroscopy 
\citep{sdssdr6}; 2: 2.5m du~Pont Telescope \citep{martini06}; 3: 
{\it Chandra X-ray Observatory} \citep{martini06,sivakoff08};
4: {\it XMM-Newton}, this work.
\label{tbl:clusterlist}
}
\end{deluxetable*}

We supplemented these observations of NORAS groups and clusters with other, 
primarily rich clusters, with redshifts in the range $0.05 < z < 0.08$. 
The details of these observations are presented in \citet{martini06} and 
\citet{sivakoff08}. Briefly, these studies are based on X-ray observations 
with {\it Chandra} and ground-based images and spectroscopy from Las 
Campanas Observatory in Chile. Further details are provided in these two 
papers. Our final list of groups and clusters is provided in 
Table~\ref{tbl:clusterlist}, along with the source of the data for each group 
or cluster. Throughout this study we separate groups and clusters based on 
whether the velocity dispersion is greater or less than 500\kms. Some of the 
implications of this choice are discussed in \S\ref{sec:xagn}. The 
division of the groups and clusters in Table~\ref{tbl:clusterlist} reflect 
this threshold value. 

\subsection{AGN Classification} \label{sec:classification}

While very luminous AGN can be unambiguously identified in almost any energy 
band, AGN become progressively more challenging to identify at lower 
luminosities when their emission may be equal or even substantially less than 
that of their host galaxy. These lower-luminosity AGN are important to 
identify to maximize the sample of AGN for demographic studies. It is also 
important to understand the completeness of the AGN selection to connect to 
other studies. At a minimum, the completeness should be expressed in terms of 
luminosity in some band, although results are more readily compared with theory 
if they can be expressed in terms of bolometric luminosity or accretion rate 
relative to the mass of the black hole. Here we identify AGN via their 
X-ray luminosity, which is estimated to represent on order 10\% of the 
bolometric luminosity with small scatter \citep[e.g.][]{elvis94,marconi04} 
and consequently it is a reasonable proxy. X-rays also have the advantage that 
they are relatively less sensitive to the effects of extinction. For the 
low-luminosity AGN we consider here, other physical processes can also produce 
comparable X-ray emission from galaxies, so in the first subsection below we 
describe our AGN classification technique in detail. The main alternative 
method to identify AGN is via visible-wavelength emission-line ratios and we 
compare our X-ray classification to this other method in the following
subsection ($\S$~\ref{sec:bptClassification}). 

In addition to careful selection of AGN via either method, characterization 
of how the AGN population varies across different environments can be 
reasonably performed with a measurement of the fraction of all galaxies of a 
given morphology that host AGN. In previous studies \citep[e.g.][]{martini02,
martini06} the AGN fraction was defined as the fraction of galaxies 
with absolute magnitude $M_R \leq -20$ mag (Vega) that host AGN with  
a broad-band [0.3-8keV] X-ray luminosity of $L_X \geq 10^{41}$ \ergs. To 
identify a comparable host luminosity range with the NORAS sample, we converted 
the SDSS extinction-corrected, $r-$band magnitudes (on the AB system) to 
Bessel $R-$band (on the Vega system) and applied a mean $k-$correction for 
each group. For both of these steps we employed the software tools described 
by \citet{blanton07}. We also adopt the evolving absolute magnitude threshold 
of $M_R \leq M_R^*(z) + 1$ introduced by \citet{martini09} to compare samples 
across a wide range in redshift, where $M_R^*(z) = M_R^*(0) - z$, and 
$M_R^*(z=0) = -21.92$. For the present sample the evolution term ($z$) is 
negligible and the main result is a second threshold approximately one 
magnitude more luminous than the previous magnitude cut. This higher threshold 
is useful because the AGN fraction increases when a higher luminosity threshold 
is used \citep{sivakoff08}. 

\subsubsection{X-ray Classification} \label{sec:xrayClassification}

We created images in the 0.5 to 8 keV band for each detector using the 
flare-cleaned event files spatially binned to give 2$''$ pixels. The images 
for the three detectors were combined to form a final image using the SAS 
task {\sc emosaic}. To identify X-ray sources, we ran the task {\sc ewavelet} 
on the merged image with a detection threshold of 5$\sigma$. The X-ray 
detections were then compared to the known members to determine matches. 
We restrict this analysis to objects within 13$'$ of the field center. All 
sources within 2$''$ of the center of a known member are considered matches; we
are motivated to use this search radius by the 1$\sigma$ positional uncertainty
of XMM of $1-2''$ \citep{watsonetal09}.
As is typical of nearby X-ray groups and poor clusters, the diffuse X-ray
emission is centered on or near the brightest galaxy in most cases 
\citep{mulchaeyzabludoff98,osmondponman04}.  
This makes searching for an AGN component difficult for the central galaxies 
and we have therefore excluded these objects from our analysis. This is 
also the case with the analysis of the richer clusters in our sample 
\citep{martini06}. 

For each member detected by {\it XMM}, we extract a surface brightness profile 
to determine the extent of the X-ray emission. We extracted source spectra in 
circular regions extending to where the surface brightness profile reaches the 
background level. Local background spectra were extracted from annular regions 
immediately surrounding the source. Using a local background of this type 
includes any additional background from the diffuse intragroup medium at the 
location of the source. Response files (RMFs and ARFs) were constructed for 
the location of the source using the SAS tasks {\sc rmfgen} and {\sc arfgen},
respectively. Source spectra were binned to have 25 counts bin$^{-1}$.

All spectral fitting was performed using {\sc XSPEC} (version 12.3). As noted 
briefly above, the four main physical processes that can produce substantial 
($L_X > 10^{40}$ \ergs) broad-band (0.3--8keV) X-ray emission from galaxies 
are AGN, a population of low-mass X-ray binaries (LMXBs), thermal emission from 
hot gas, and emission from the high-mass X-ray binaries (HMXBs) and supernova 
remnants associated with substantial recent star formation. We classify an 
X-ray source as an AGN if the observed X-ray luminosity exceeds that expected 
from these other physical processes that can produce X-ray emission. Our 
basic procedure is described in \citet{sivakoff08}, which improves on the 
earlier procedure employed by \citet{martini06}. Briefly, we use relations 
between X-ray luminosity, $K-$band luminosity, and star formation rate to 
determine the expected contribution from LMXBs \citep[cf.\ ][]{kim04}, star 
formation \citep[cf.\ ][]{grimm03}, and halos of hot gas \citep[cf.\ ][]{sun07} 
and classify a galaxy as an AGN if the X-ray luminosity exceeds the expected 
contribution from these other sources of emission. 

Because these {\it XMM} observations often have sufficient counts for spectral 
fits, we fit two spectral models to better classify AGN when the data are 
sufficient. These models are a single power-law component to represent the 
combined emission of the LMXBs and any AGN component and a thermal 
component to represent any emission from hot gas. We then estimate the X-ray 
binary emission expected from the $K-$band luminosity of the galaxy and
consider any excess emission from the power-law component to be due to
an AGN. In all our fits the neutral hydrogen column density
is fixed at the Galactic value given in \citet{kalberla05}. We also fix the 
power-law index ($\Gamma$) to 1.7. For the thermal component, we use the 
{\sc MEKAL} model in {\sc XSPEC} with the abundance fixed at 0.8 solar. 
We simultaneously fit the spectra from all three EPIC detectors. For galaxies 
with at least several hundred counts, it is usually possible to constrain 
both the thermal and powerlaw components. In some cases, only one component is 
required to produce an adequate fit (i.e. the normalization of the second 
component is consistent with zero). For galaxies with a small number of 
counts, it is not possible to distinguish between the possible spectral
models. For these objects, we have estimated the X-ray luminosity assuming a 
power-law model alone. The resulting luminosities of the power-law and thermal 
components for each galaxy are given in Table~\ref{tbl:xdata}. We note that if 
a thermal model is assumed for the cases where the spectral model cannot be 
determined, the resulting luminosities would be lower by a factor of 
approximately two.

\begin{deluxetable*}{lcclccl}
\tabletypesize{\scriptsize}
\tablecaption{X-ray Detected Group Members}
\tablehead{
    \colhead{Galaxy} &  
    \colhead{M$_R$} &
    \colhead{L$_{\rm K}$} & 
    \colhead{Model} & 
    \colhead{L$_{\rm X}$$_{\rm powerlaw}$} & 
    \colhead{L$_{\rm X}$$_{\rm thermal}$} &
    \colhead{Class} \\ 
\colhead{} & \colhead{} & \colhead{} & \colhead{} & \colhead{10$^{40}$ erg s$^{-1}$} & \colhead{10$^{40}$ erg s$^{-1}$} \\
\colhead{} & \colhead{} & \colhead{} & \colhead{} & \colhead{(0.3-8.0 keV)} & \colhead{(0.3-8.0 keV)} & \colhead{} \\
 &  &  &  &  &  & \\                                                                                    
\colhead{(1)} &
\colhead{(2)} &
\colhead{(3)} &
\colhead{(4)} &
\colhead{(5)} &
\colhead{(6)} &
\colhead{(7)}
}
\startdata
2MASXJ07463295+3101213      & -22.6   &  11.42   & Po$^1$  & 18.6$^{+10.3}_{-9.1}$   & -                        & AGN      \\         
2MASXJ07462331+3101183      & -22.2   &  11.36   & Po$^1$  & 10.3$^{+7.4}_{-5.5}$    & -                        & inactive \\    
2MASXJ08445063+4302479$^*$  & -23.1   &  11.67   & Po      & 24.1$^{+9.0}_{-8.5}$    & -                        & AGN  \\         
2MASXJ10230356+3838176      & -21.0   &  10.70   & Po$^1$  & 72.2$^{+31.4}_{-30.8}$  & -                        & AGN      \\         
2MASXJ10223745+3834447      & -23.2   &  11.69   & Po+ Th  & 104.9$^{+27.8}_{-39.1}$ & 33.0$^{+48.5}_{-8.5}$    & AGN      \\         
2MASXJ10220069+3829145      & -21.9   &  11.03   & Po$^1$  & 23.2$^{+8.8}_{-8.2}$    & -                        & AGN      \\         
2MASXJ11231618+6706308      & -22.1   &  11.22   & Po$^1$  & 16.6$^{+10.7}_{-7.8}$   & -                        & AGN      \\         
2MASXJ11221610+6711219      & -21.5   &  10.96   & Po      & 14.0$^{+7.3}_{-6.5}$    & -                        & AGN      \\         
2MASXJ11223691+6710171$^n$  & -21.6   &  11.03   & Po$^1$  & 8.5$^{+13.9}_{-4.2}$    & -                        & AGN      \\         
SDSSJ112333.56+671109.9     & -20.1   &    -     & Po+ Th  & 40.7$^{+15.1}_{-12.0}$  & 14.9$^{+18.1}_{-4.5}$    & AGN      \\         
2MASXJ12043806+0147156      & -22.5   &  11.33   & Po+ Th  & 7.2$^{+3.7}_{-2.0}$     & 2.8$^{+4.3}_{-1.2}$      & inactive \\    
2MASXJ12225772+1032540      & -21.3   &  11.08   & Po+ Th  & 2.6$^{+2.1}_{-1.1}$     & 1.4$^{+2.7}_{-0.4}$      & inactive \\    
2MASXJ13242889+1405332      & -22.3   &  11.43   & Po      & 10.0$^{+1.8}_{-1.8}$    & -                        & inactive \\    
2MASXJ14403793+0322375      & -22.3   &  11.38   & Po      & 6.1$^{+1.7}_{-1.9}$     & -                        & inactive 
\enddata
\tablecomments{
X-ray measurements and classifications. Columns are: (1) Galaxy name; (2) Host 
galaxy $R-$band absolute magnitude; (3) Total $K-$band luminosity from 2MASS 
(for SDSSJ112333.56+671109.9 $K=14$ mag was assumed based on the colors of 
other group members); (4) Model fit to the X-ray data where Po is a power-law 
fit, Th is a thermal model, and Po$^1$ indicates a power-law was assumed 
(see $\S$~\ref{sec:xrayClassification}); (5) X-ray luminosity of the power law 
component; (6) X-ray luminosity of the thermal component; (7) classification of 
the galaxy as either an AGN or as inactive. $L_{X,po}$ and $L_{X,th}$ are 
in units of $10^{40}$ \ergs\ and are broad-band (0.3-8keV) measurements. 
The $^*$ superscript in Column~1 denotes the single galaxy in the NORAS sample that 
is classified as an AGN based both on its X-ray properties and its emission 
lines. The $^n$ superscript refers to an AGN below our $L_X = 10^{41}$ \ergs\ limit
and thus not included in the sample statistics. 
\label{tbl:xdata}
}
\end{deluxetable*}

\begin{figure}
\epsscale{1.0}
\plotone{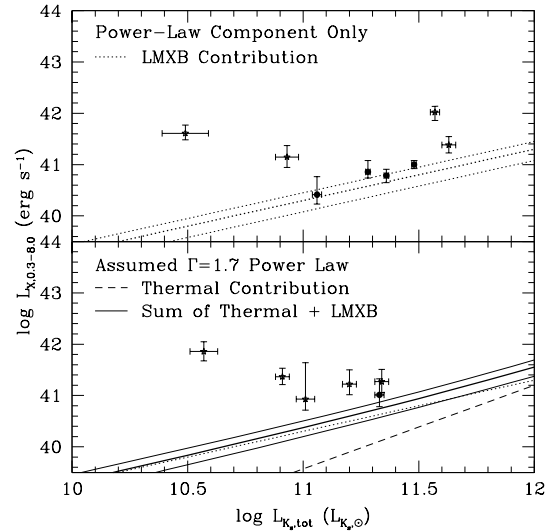}
\caption{The broad-band X-ray luminosity $L_{X,B}$ vs.\ the near-IR luminosity, 
$L_{K_s,\mathrm{tot}}$ for the galaxies displayed in Table~\ref{tbl:xdata}. 
The top panel shows the broad-band X-ray luminosity $L_{X,B}$ of the 
best-fit power-law component for all of these eight galaxies 
compared to the $K-$band luminosity. 
Here we compare to the expected relation for LMXBs ({\it dotted line}) from 
\citet{kim04} and classify four of eight galaxies as X-ray AGN ({\it stars}). 
The bottom panel shows the relation for sources with insufficient counts for 
spectral modeling. Here we assume a $\Gamma = 1.7$ power law and compare with 
the sum ({\it solid line}) of the LMXB relation and a thermal model ({\it 
dashed line}) from \citet{sun07}. Five of the sources are classified as 
AGN, although one is below $10^{41}$ \ergs\ and is not included in the 
statistical analysis. In both panels the thicker line is the relation and the 
thinner lines are $\pm 1\sigma$ uncertainties. Other galaxies are classified 
as inactive ({\it filled circles}, see Table~\ref{tbl:xdata}). 
\label{fig:lxlk}}
\end{figure}

The additional spectral information available for many of these sources better 
constrains the nature of the X-ray emission. Specifically, eight of the 14 
X-ray sources associated with members have sufficient signal-to-noise ratio 
to determine if the X-ray emission was best-fit by a power-law model, thermal 
model, or both. This improves the classification over, e.g. \citet{sivakoff08} 
as we can then compare the luminosity of the best-fit power-law model to just 
the expected emission from LMXBs, HMXBs, and an AGN component, and exclude 
the thermal model because of its different spectrum. The top panel of 
Figure~\ref{fig:lxlk} shows the broad-band X-ray luminosity of the 
best-fit power-law component for all of these eight galaxies 
compared to the $K-$band luminosity. 
Note that whereas in \citet{sivakoff08} we used the 2MASS $K_{20}$ magnitude, 
here we employ the $K_{total}$ magnitude as this is a better match to the 
aperture used for the X-ray photometry. The figure also shows the expected 
relationship between X-ray luminosity and $K-$band luminosity for LMXBs from 
\citet{kim04}, where the thicker line is the relation and the thinner lines 
are $\pm 1\sigma$ uncertainties. Four of the eight X-ray sources fall on the 
LMXB relation and we classify these galaxies as inactive (see 
Table~\ref{tbl:xdata}). The remaining four are at least $2\sigma$ more X-ray 
luminous than would be expected from LMXBs alone and we therefore classify 
these galaxies as X-ray AGN. We note that these classifications are the same 
as we would have assigned based on our previous approach with an 
$L_X = 10^{41}$ \ergs\ threshold \citep{martini06,sivakoff08}. 

The other six galaxies had sufficiently faint X-ray emission that we were 
unable to accurately model their X-ray spectra. In these cases we followed 
the procedure used by \citet{sivakoff08} and assumed a $\Gamma = 1.7$ 
model and measured the X-ray luminosity of that model. These six sources 
are shown in the bottom panel of Figure~\ref{fig:lxlk} along with the 
same LMXB relation shown in the top panel, a relation between $L_X$ and 
$L_K$ for thermal emission from hot gas adapted from \citet{sun07}, and the 
sum of these two relations, where again the inactive galaxy relation is 
represented by the thicker line and the $\pm 1\sigma$ uncertainties are 
represented by thinner lines. The hot gas relation is modified from that 
presented by \citet{sun07} because their measurements were in the soft band 
(0.5-2keV) and ours are broad-band (0.3-8keV) measurements. We therefore 
multiplied their X-ray 
luminosity by the flux ratio of a $kT = 0.7$keV thermal bremsstrahlung 
model in the broad band and soft band. In practice none of the sources 
in the lower panel are sufficiently luminous in the $K-$band that we expect 
a substantial thermal component. Five of the six sources are above the 
relation by at least $2\sigma$ and we classify these galaxies as X-ray 
AGN, although one of these is not included in our statistical analysis because 
it has $L_X < 10^{41}$ \ergs. We also note that none of these 14 galaxies 
appears to have sufficient 
star formation to contribute significantly to the X-ray luminosity. To 
check this we estimated star formation rates for each galaxy from the 
H$\alpha$ flux \citep[e.g.][]{kennicutt98,brinchmann04} and then calculated 
the expected X-ray luminosity from \citet{grimm03}. In all cases the estimated 
X-ray luminosity due to star formation was at least an order of magnitude below 
the observed luminosity. For the five groups and clusters in common with 
\citet{sivakoff08} we retain the same AGN classifications presented in that 
paper. 

\subsubsection{Emission-Line Classification} \label{sec:bptClassification}

The traditional technique to identify low-luminosity AGN in emission-line 
galaxies is through use of line ratios such as \OIII\ and \NII\ on a ``BPT 
Diagram'' \citep{bpt81}. We have measurements of these four emission lines for 
most (349) of the galaxies in the sample from the NORAS catalog from the 
MPA-JHU galaxy catalog for the SDSS Data Release~7\footnote{http://www.mpa-garching.mpg.de/SDSS/DR7/}. We use the 
emission-line measurements and errors calculated by the MPA-JHU group, which 
have been corrected for stellar absorption, and then calculate emission line 
ratios. If the line flux measurements for \NII\ are larger than 3 times the 
error associated with the measurements, we keep the data for analysis. We are 
less conservative for the \OIII\ ratio because galaxies with precise \NII\ 
measurements can be unambiguously identified as AGN even if the \OIII\ 
measurements are less certain \citep[e.g.][]{shen07}. In addition, edge-on 
galaxies might obscure the bluer $[\mathrm{O III}]$ and $\mathrm{H}\beta$ 
lines, leading to a misclassification. We hope to keep these possibly 
obscured but still unambiguous AGN in our 
sample by being less stringent with our condition on the \OIII\ measurement 
errors. 
\begin{figure}
\epsscale{1.0}
\plotone{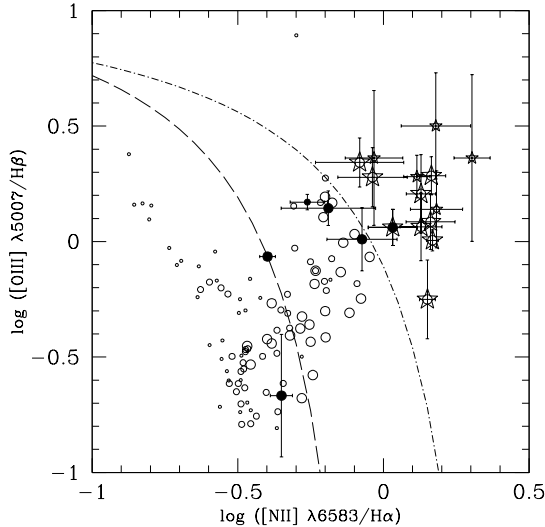}
\caption{A BPT Diagram \citep{bpt81} for all galaxies with sufficiently bright 
emission-lines in our NORAS sample. Inactive galaxies ({\it open circles}) are 
separated based on whether $M_R > -20$ mag ({\it small open circles}), $-20 
\ge M_R \ge M_R^*+1$ ({\it medium open circles}), or have $M_R \leq M_R^*+1$
({\it large open circles}). The X-ray AGN ({\it filled circles}) and 
BPT AGN ({\it stars}) are similarly scaled based on host galaxy luminosity and 
all AGN have error bars. Galaxies are classified as BPT AGN if they fall above 
the \citet{kewley01} criterion ({\it dot-dashed line}). The criterion of 
\citet{kauffmann03} is also shown ({\it dashed line}).  
\label{fig:bpt}}
\end{figure}

Of the 349 galaxies in the MPA-JHU galaxy catalog, 116 have sufficiently 
bright spectral lines to place on the BPT Diagram shown in 
Figure~\ref{fig:bpt}. This figure illustrates the values of \NII\ and \OIII\ 
for these 116 galaxies. We identify galaxies as ``BPT AGN'' if these two line 
ratios place them above the threshold suggested by \citet{kewley01} as the 
maximum limit that could be produced by extreme starburst activity. This is a 
very conservative limit and many confirmed AGN do not meet this criterion. For 
reference, we also show the less conservative threshold of \citet{kauffmann03}. 
Of the 116 galaxies with sufficient emission-line detections to fall on this 
diagram, 14 were identified as BPT AGN. A total of 199 out of the 349 galaxies 
with SDSS spectra also have $M_R \leq -20$ mag, so the BPT AGN fraction for 
this sample is $14/199 = 0.070^{+0.024}_{-0.019}$ (all BPT AGN have $M_R \leq
-20$ mag). For comparison, six of these 199 galaxies or $\sim 3$\% were 
identified as X-ray AGN (note that not all X-ray AGN had sufficient 
spectral-line data). It is also striking that only one galaxy is 
classified as both an X-ray AGN and a BPT AGN. The samples are nearly disjoint. 
Further information about the galaxies identified as BPT AGN is provided in 
Table~\ref{tbl:BPTdata}.

\begin{deluxetable*}{lccc}
\tabletypesize{\scriptsize}
\tablecaption{BPT AGN}
\tablehead{
    \colhead{Galaxy} &  
    \colhead{M$_R$} &
    \colhead{log(\OIII)} &
    \colhead{log(\NII)} \\
 &  &  &  \\
\colhead{(1)} &
\colhead{(2)} &
\colhead{(3)} &
\colhead{(4)}
}
\startdata
   2MASXJ01095902+1358155               & -20.8  &  0.182  $\pm$  0.090 &  0.140  $\pm$  0.166  \\
  SDSSJ010957.88+140320.1               & -20.4  & -0.032  $\pm$  0.099 &  0.362  $\pm$  0.292  \\
  SDSSJ011021.57+135421.4               & -20.2  &  0.180  $\pm$  0.119 &  0.501  $\pm$  0.230  \\
  2MASXJ07462331+3101183                & -22.2  &  0.164  $\pm$  0.049 &  0.285  $\pm$  0.082  \\  
   2MASXJ07470054+3058205               & -21.7  &  0.129  $\pm$  0.050 &  0.207  $\pm$  0.170  \\
   2MASXJ08445063+4302479 $^*$          & -23.1  &  0.032  $\pm$  0.084 &  0.061  $\pm$  0.078  \\
  SDSSJ100311.10+323511.3               & -20.2  &  0.304  $\pm$  0.062 &  0.362  $\pm$  0.361  \\
   2MASXJ10213991+3831195               & -21.5  & -0.038  $\pm$  0.119 &  0.278  $\pm$  0.128  \\
   2MASXJ11223691+6710171 (VIIZw392)    & -22.4  &  0.168  $\pm$  0.020 &  0.005  $\pm$  0.046  \\
  SDSSJ112425.38+671940.0               & -20.3  &  0.115  $\pm$  0.045 &  0.281  $\pm$  0.092  \\
   2MASXJ12041899+015054  (CGCG013-058) & -21.2  &  0.129  $\pm$  0.072 &  0.064  $\pm$  0.146  \\
   2MASXJ12230667+1037170 (NGC4325)     & -22.5  &  0.151  $\pm$  0.027 & -0.251  $\pm$  0.171  \\
   2MASXJ12225772+1032540               & -21.3  & -0.082  $\pm$  0.152 &  0.343  $\pm$  0.106  \\
   2MASXJ13241000+1358351 (NGC5129)     & -23.1  &  0.162  $\pm$  0.084 &  0.086  $\pm$  0.120
\enddata
\tablecomments{
Galaxies identified as AGN on the BPT Diagram \citep{bpt81} shown in 
Figure~\ref{fig:bpt}. Columns are: (1) The name (alternate name) of the galaxy; 
(2) $M_R$; (3) calculated value of log \OIII\ from the MPA-JHU database; 
(4) calculated value of log \NII\ from the MPA-JHU database. The $^*$ superscript in 
Column~1 denotes the only galaxy that is classified as an AGN based both on its 
X-ray properties and its visible-wavelength emission lines.  
\label{tbl:BPTdata}
}
\end{deluxetable*}

\section{Morphological Analysis} \label{sec:morphology}

Though galaxy morphology has been used extensively to study galaxy evolution, 
until recently this property was determined by eye. This process, though 
largely repeatable, lacks quantitative robustness and is a protracted process 
for large numbers of objects. The alternative---measuring morphology in an 
automated fashion---is not simple to implement. Only in the last decade, with 
the development of large-format, linear detectors and substantial computational 
resources, has it become commonplace to classify galaxies using quantitative 
and repeatable techniques. Datasets consisting of large numbers of galaxies 
have made visual inspection intractable as a method for determining morphology 
\citep[although there have been novel approaches for morphological 
identification via visual inspection; see][]{lintottetal08}, while abundant
computational resources have made large-scale, quantitative analyses more 
feasible. 

Various methods to ascertain morphology quantitatively exist in the literature 
\citep{conselice00,simard02,goto03,lotz04}, and we chose to use the galaxy 
fitting code GALFIT \citep{peng02} to measure the morphological properties 
for galaxies in our sample. One of our main motivations for the choice of 
GALFIT was the work of \citet{haeussler07}, who compared GALFIT to GIM2D 
\citep{simard02} and concluded that GALFIT has advantages in its ability to 
simultaneously fit neighboring galaxies in a crowded field and benefits from a 
dramatic increase in execution speed. GALFIT is designed to extract structural 
components from galaxies by fitting two-dimensional light profiles with an 
arbitrary number of parametric functions that are suitable for describing the 
surface brightness distribution of galaxies. Although the code was authored 
to fit 
subtle structures of well-resolved galaxies with many-component models 
simultaneously, it is also effective at handling large numbers of galaxies 
imaged at lower resolution by fitting their surface brightness profiles with 
relatively simple models. We utilize the latter capability in our analysis. 
GALFIT takes as input a simple text file and is very customizable, allowing 
easy extension via a wrapper script. A final advantage is that GALFIT can use 
a variety of analytic functions singly or simultaneously, including the
S\'ersic profile. In the next subsections we describe the fitting procedure 
in more detail, including how we parametrize galaxy morphology, and 
present the results of our fits of the X-ray and BPT AGN. We fit models to 
all of the galaxies in our sample that have imaging data.

\subsection{Model Fits}

We have adopted two parameters to quantify galaxy morphology: the value 
of the S\'ersic index $n$ and the bulge-to-total flux ratio $B/T$. The 
S\'ersic profile is defined to be: 
 \begin{equation}
 \Sigma(r) = \Sigma_{e} \mathrm{exp} \left[ -\kappa \left( 
       \left( \frac{r}{r_{e}} \right)^{1/n} -1 \right) \right] ,
 \end{equation}
where $\Sigma_e$ is the surface brightness (flux per unit area) at the 
effective radius $r_e$, $r_e$ is the half-light radius, and the 
eponymous index $n$ characterizes the shape of the light profile. The parameter 
$\kappa$ is set by the constraint that $r_e$ is the half-light radius. 
The S\'ersic index $n$ has been commonly used to separate early-type and 
late-type galaxies in the literature. Studies based on SDSS imaging data find 
that $n = 2.5$ is a reasonable point to distinguish these two types 
\citep[e.g.][]{bell04,mcintosh05} and we too adopt $n\geq2.5$ to 
identify early-type galaxies. 

The bulge-to-total flux ratio is measured from a classic bulge-disk 
decomposition. Here we fit a bulge component with a \citet{devauc48} 
$r^{1/4}$ profile and a disk component with an exponential surface 
brightness profile. The fraction of the total flux in the bulge component
relative to the total (bulge + disk) flux then provides the ratio $B/T$. 
Throughout this work we will primarily rely on the S\'ersic index to 
classify galaxies, but we will use the $B/T$ as a consistency check on 
our results. Note also that $n = 4$ is equivalent to the $r^{1/4}$
profile and $n = 1$ is equivalent to an exponential disk profile. 

When available we also compared our calculated morphological parameters to 
the SDSS \texttt{fracDeV} parameter, which is calculated by the SDSS pipeline 
and serves as another quantitative measure of galaxy morphology. The 
\texttt{fracDeV} parameter is very similar to our bulge-to-total flux 
decomposition. It is obtained by fitting the surface brightness profile of a 
galaxy with exponential and de~Vaucouleurs components and then keeping the 
fractional contribution of the latter. 
\citet{bernardi05} identified \texttt{fracDeV}\ $ > 0.8$ to identify early-type 
galaxies. 

As noted above, GALFIT accepts an input text file that specifies the initial 
conditions for models and other options and parameters related to the fit. 
GALFIT can accommodate as many independent models as the user desires, limited 
only by computational resources. It is also possible to simultaneously fit 
adjacent or blended objects in addition to the target to remove potential 
contamination and obtain a more robust fit. GALFIT convolves the model with a 
point spread function (PSF) supplied by the user, subtracts the convolved 
model from the input image, and computes the reduced chi squared, \chisqnu:
  \begin{equation}
  \chi^{2}_{\nu} = \frac{1}{N_{\mathrm{dof}}} \sum_{x=1}^{nx} \sum_{y=1}^{ny} 
                   \frac{ \left( \mathrm{flux}_{x,y} - 
                   \mathrm{model}_{x,y} \right) ^{2} }
                    {\sigma_{x,y}^{2}},
  \end{equation}
where $N_{\mathrm{dof}}$ is the number of degrees of freedom in the model, flux 
and model are the pixel values of the original image and analytic model, 
respectively, and $\sigma_{x,y}^{2}$ is the error in each pixel. GALFIT 
minimizes \chisqnu\ using a Levenberg-Marquardt algorithm, a downhill-gradient 
type algorithm suited for searching large parameter spaces quickly. Additional 
and optional input includes a bad pixel map specifying which pixels should be 
excluded from the \chisqnu\ calculation (i.e. masked out) and initial guesses 
for the many free parameters, including the astrometric and morphological 
quantities of the target. 

GALFIT is useful in its extensibility and we took advantage of this by creating 
a wrapper script in Python and an algorithm to fit many target objects with 
little to no user interaction. The results are comparable to a human user 
fine tuning GALFIT input parameters until an ideal fit is obtained. 
The input to this process is a list of astrometric coordinates of targets 
and the FITS images in which these targets are imaged. For each target the 
script determines if it lies within the boundaries of a given image and 
obtains initial morphological parameters (e.g. a measure of the galaxy's 
radius, magnitude, ellipticity, object position, and position angle) 
with \textsc{SExtractor} \citep{bertin-arnouts96}. At this stage we 
retain information about all detected objects within some arbitrary number 
(found using trial and error) of effective radii from the target. The fitting 
region supplied to GALFIT is determined in a similar way. Based on the 
parameters of the objects in the field of view, we either add the pixels 
associated with the object to the bad pixel file, masking them out and 
removing them from the \chisqnu\ calculation and fitting procedure, or fit the 
object in addition to the target. This discriminatory algorithm compares 
object brightness to the target brightness (e.g. an object very much dimmer 
than the target will likely be masked out rather than fit) and the distance
from the target (e.g. an object separated by many target galaxy-radii  
will likely be masked out rather than fit). In this way we only 
fit additional objects if they are likely to contaminate the fit of our
target, and thus its morphological properties. Fitting a superfluous
number of objects is computationally wasteful and complicates finding
a unique minimum in \chisqnu\ space.  

Our wrapper script generates a GALFIT input file for both a single component 
S\'ersic profile fit and a two component S\'ersic profile $+$ Exponential Disk 
fit. Several iterations of this dual method fitting occur if the resultant 
\chisqnu\ values for the two parameterizations differ by more than a small 
amount, with the previous fit parameters used for subsequent iterations, to 
insure that the solution obtained is not the result of the minimization 
algorithm getting lost in a local minimum. This propensity to get lost in 
a local minimum and the question of the uniqueness of a multi-component 
solution in large parameters spaces is an issue for algorithms like GALFIT, 
and \citet{peng02} addresses this question in depth (see their $\S$~3.3).

Another requirement for GALFIT is an estimate of the PSF shape. These were 
created manually for each image by approximating them with a Gaussian 
generated with the standard \texttt{gauss} task in IRAF. Effective radii of 
the PSFs were estimated by examining several stars in each image. We determined 
that small variations in the radius of the PSF negligibly affected the 
parameters of the model, confirming the robustness of this approximate 
method of PSF creation. 

When completed, the wrapper saves: 1) a postage stamp sized image of the 
original galaxy (i.e. a cropped section of the wide-field FITS image where the 
target fills the frame); 2) an image of the same size that displays the model 
generated by GALFIT convolved with the user-defined PSF; 3) a residual image: 
the original data less the convolved model. This third image should, if the 
model is perfect, show a scene identical to the input image, with the 
exception that the region of sky previously occupied by the target galaxy 
should be indistinguishable from noise (although somewhat increased noise due 
to the contribution of the subtracted object). Unsurprisingly, this ideal 
case is rare (though not nonexistent). These three images are saved for each 
fit technique (single and two-fit methods) for quick visual inspection. We also 
retain the fit log files generated by GALFIT, FITS images that contain the 
GALFIT models generated by the algorithm, and a text file with the relevant 
morphological parameters and object information for later use. The script that 
automates these tasks works for an arbitrarily large set of input targets.  

\subsection{Fit Results} 

\begin{figure*}
\epsscale{0.64}
\plotone{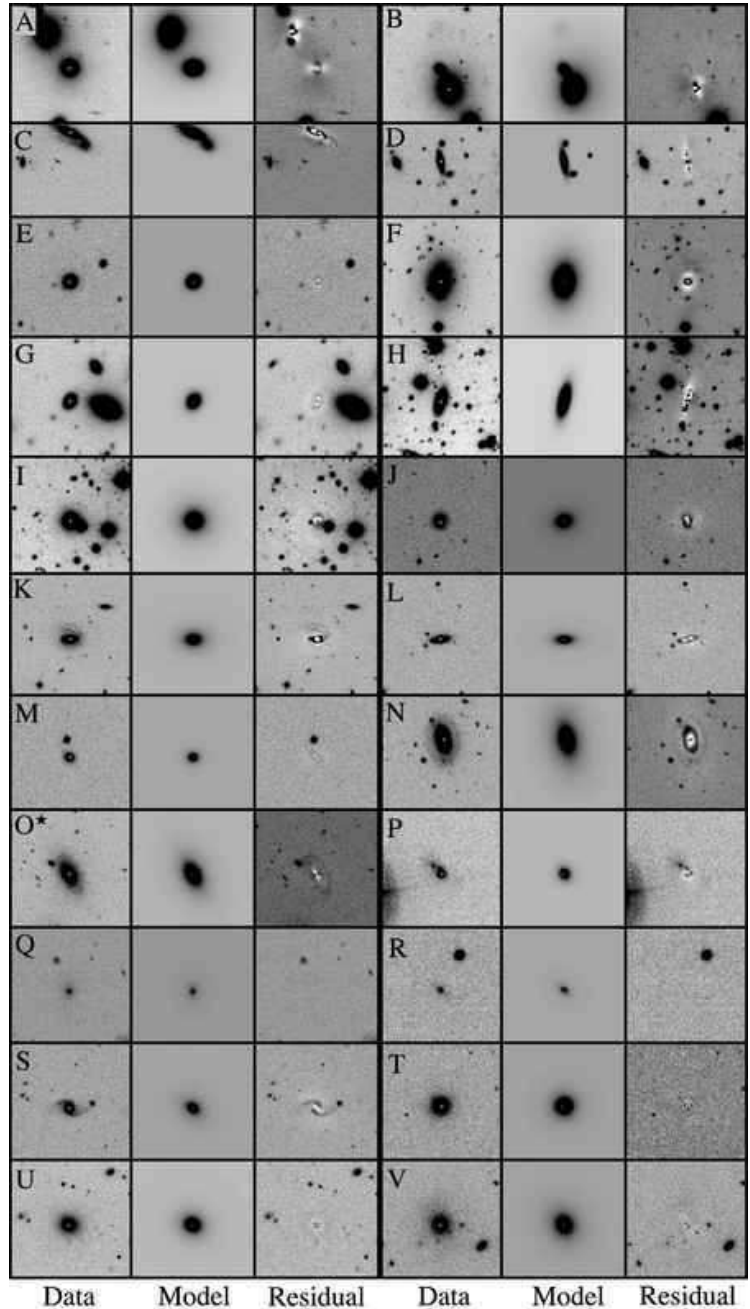}
\caption{
Data, GALFIT models, and residuals of all the X-ray AGN for our entire sample. 
For details of the AGN identification see \S\ref{sec:xrayClassification}, 
\citet{martini06}, and \citet{sivakoff08}. 
The morphological parameters associated with these fits are listed in 
Table~\ref{tbl:agnExamples}. The star symbol, next to Panel~O, denotes the only 
galaxy in our sample identified as both a BPT AGN and an X-ray AGN. 
This object is also in Panel~F in Figure~\ref{fig:BPTagnExamples}.
\label{fig:agnExamples}}
\end{figure*}

\begin{figure*}
\epsscale{0.64}
\plotone{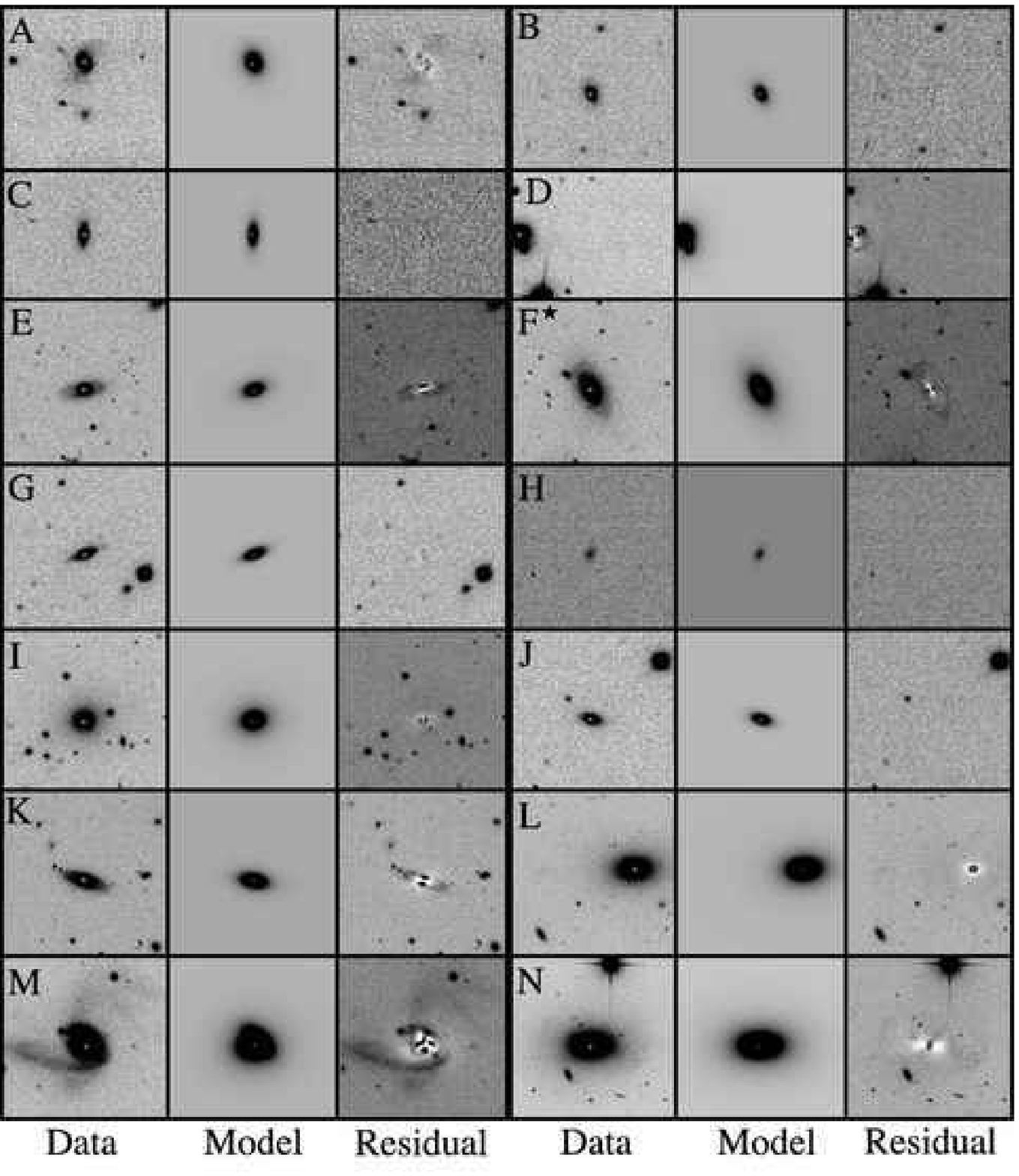}
\caption{
Data, GALFIT models, and residuals of all the BPT AGN for our entire sample. 
For details of AGN identification see $\S$~\ref{sec:bptClassification}.
The morphological parameters associated with these fits are listed in 
Table~\ref{tbl:BPTagnExamples}. The star symbol, next to Panel~F, denotes the 
only galaxy in our sample identified as both a BPT AGN and an X-ray AGN. 
This object is also in Panel~O in Figure~\ref{fig:agnExamples}.
\label{fig:BPTagnExamples}}
\end{figure*}

The three-image results for all of the X-ray AGN are shown in 
Figure~\ref{fig:agnExamples}, while the results for the BPT AGN are shown in 
Figure~\ref{fig:BPTagnExamples}. The model parameters associated with all of 
these fits are listed in Tables~\ref{tbl:agnExamples} and 
\ref{tbl:BPTagnExamples}. Only one object falls in both samples and 
consequently appears in both figures and tables. While these figures 
demonstrate our results on AGN, described further below in \S\ref{sec:bptagn}, 
they also are fairly representative of our morphological fits to the inactive 
group and cluster galaxies. 

\begin{deluxetable*}{cllcccc}
\tablecolumns{7}
\tabletypesize{\scriptsize}
\tablecaption{Fit Properties from Figure~\ref{fig:agnExamples} --- X-ray AGN}
\tablehead{
\colhead{Panel} &
\colhead{Cluster/Group} &
\colhead{Object Name} &
\colhead{S\'ersic Index} &
\colhead{\chisqnu} &
\colhead{f$_{\mathrm{bulge}}$ / f$_{\mathrm{total}}$} &
\colhead{\chisqnu} \\
 &  &  &  &  &  &  \\                                                                                    
\colhead{(1)} &
\colhead{(2)} &
\colhead{(3)} &  
\colhead{(4)} &
\colhead{(5)} &
\colhead{(6)} &
\colhead{(7)}
}
\startdata
A     & A3125           & J032723.4-532535.5                          & 3.18 & 1.965 & 0.930 & 1.875 \\ 
B     & A3125           & J032725.3-532506.6                          & 2.27 & 1.635 & 0.768 & 1.638 \\ 
C     & A3125           & J032705.1-532140.9                          & 2.63 & 1.739 & 0.724 & 1.630 \\ 
D     & A3128           & J033039.2-523205.7                          & 1.24 & 1.300 & 0.109 & 1.236 \\ 
E     & A3128           & J032941.4-522935.7                          & 1.36 & 1.182 & 0.335 & 1.178 \\ 
F     & A3128           & J033051.0-523031.2                          & 4.53 & 1.528 & 0.964 & 1.652 \\ 
G     & A3128           & J033017.3-523408.9                          & 1.75 & 1.585 & 0.458 & 1.576 \\ 
H     & A644            & J081748.1-073731.7                          & 2.16 & 9.746 & 0.575 & 7.099 \\ 
I     & A644            & J081739.5-073309.0                          & 12.0 & 10.58 & 0.371 & 7.762 \\ 
J     & A85             & J004311.6-093816.1                          & 3.06 & 1.583 & 0.400 & 1.568 \\ 
K     & A85             & J004130.3-091545.9                          & 3.57 & 1.712 & 0.085 & 1.606 \\ 
L     & A89B            & J004314.1-092145.2                          & 2.89 & 1.538 & 0.545 & 1.532 \\ 
M     & RXCJ1122.2+6712 & SDSSJ12333.56+671109.9                      & 1.56 & 1.498 & 0.053 & 1.495 \\
N     & RXCJ0746.6+3100 & 2MASXJ07463295+3101213                      & 4.28 & 2.044 & 1.000 & 2.038 \\
O$^*$ & RXCJ0844.9+4258 & 2MASXJ08445063+4302479 (CGCG208-041)        & 4.59 & 1.551 & 0.812 & 1.552 \\
P     & RXCJ1022.0+3830 & 2MASXJ10220069+3829145                      & 2.37 & 6.604 & 0.808 & 6.556 \\
Q     & RXCJ1022.0+3830 & 2MASXJ10223745+3834447 (NGC 3219)           & 3.49 & 6.268 & 0.851 & 6.268 \\
R     & RXCJ1022.0+3830 & 2MASXJ10230356+3838176                      & 9.38 & 6.672 & 0.534 & 6.672 \\
S     & RXCJ1122.2+6712 & 2MASXJ11221610+6711219                      & 6.88 & 1.625 & 0.355 & 1.587 \\
T$^n$ & RXCJ1122.2+6712 & 2MASXJ11223691+6710171 (VIIZw394)           & 3.35 & 1.550 & 0.894 & 1.547 \\
U     & RXCJ1122.2+6712 & 2MASXJ11231618+6706308                      & 4.60 & 1.520 & 1.000 & 1.524 \\
V     & A89B            & J004300.63-091346.4                         & 5.97 & 1.544 & 0.999 & 1.569
\enddata
\tablecomments{GALFIT output parameters from fits to all X-ray AGN. The $^*$ 
superscript in Column~1 identifies the only galaxy in our sample that is classified 
as both a BPT AGN and an X-ray AGN. This object is Object~F in 
Table~\ref{tbl:BPTagnExamples}. The $^n$ superscript refers to an AGN below 
our $L_X = 10^{41}$ \ergs\ limit. 
\label{tbl:agnExamples}
}
\end{deluxetable*}

\begin{deluxetable*}{cllcccc}
\tablecolumns{7}
\tabletypesize{\scriptsize}
\tablecaption{Fit Properties from Figure~\ref{fig:BPTagnExamples} --- BPT AGN}
\tablehead{
\colhead{Panel} &
\colhead{Cluster/Group} &
\colhead{Object Name} &
\colhead{S\'ersic Index} &
\colhead{\chisqnu} &
\colhead{f$_{\mathrm{bulge}}$ / f$_{\mathrm{total}}$} &
\colhead{\chisqnu} \\
 &  &  &  &  &  &  \\                                                                                    
\colhead{(1)} &
\colhead{(2)} &
\colhead{(3)} &  
\colhead{(4)} &
\colhead{(5)} &
\colhead{(6)} &
\colhead{(7)} 
}
\startdata
A     & RXCJ0110.0+1358  &  2MASXJ01095902+1358155                     & 2.99 & 1.625 & 0.932 & 1.648  \\
B     & RXCJ0110.0+1358  & SDSSJ010957.88+140320.1                     & 2.20 & 1.567 & 0.398 & 1.564  \\
C     & RXCJ0110.0+1358  & SDSSJ011021.57+135421.4                     & 1.96 & 1.523 & 0.562 & 1.524  \\
D     & RXCJ0746.6+3100  &  2MASXJ07462331+3101183                     & 2.55 & 2.221 & 0.509 & 2.112  \\
E     & RXCJ0746.6+3100  &  2MASXJ07470054+3058205                     & 2.96 & 1.604 & 0.613 & 1.582  \\
F$^*$ & RXCJ0844.9+4258  &  2MASXJ08445063+4302479 (CGCG208-041)       & 4.59 & 1.551 & 0.812 & 1.552  \\
G     & RXCJ1002.6+3241  & SDSSJ100311.10+323511.3                     & 1.51 & 1.424 & 0.472 & 1.427  \\
H     & RXCJ1022.0+3830  &  2MASXJ10213991+3831195                     & 1.13 & 6.372 & 0.022 & 6.372  \\
I     & RXCJ1122.2+6712  &  2MASXJ11221537+671318  (VIIZw392)          & 4.86 & 1.595 & 0.667 & 1.582  \\
J     & RXCJ1122.2+6712  & SDSSJ112425.38+671940.0                     & 1.69 & 1.531 & 0.509 & 1.532  \\
K     & RXCJ1204.4+0154  &  2MASXJ12041899+015054  (CGCG013-058)       & 2.17 & 1.798 & 0.726 & 1.747  \\
L     & RXCJ1223.1+1037  &  2MASXJ12230667+1037170 (NGC4325)           & 2.42 & 1.627 & 0.884 & 1.579  \\
M     & RXCJ1223.1+1037  &  2MASXJ12225772+1032540                     & 3.96 & 2.313 & 0.977 & 2.316  \\
N     & RXCJ1324.1+1358  &  2MASXJ13241000+1358351 (NGC5129)           & 4.22 & 1.805 & 0.792 & 1.762
\enddata
\tablecomments{GALFIT output parameters from fits to all BPT AGN. 
The $^*$ superscript in Column~1 identifies the only galaxy in our sample that is 
classified as both a BPT AGN and an X-ray AGN. This object is Object~O in
Table~\ref{tbl:agnExamples}.
\label{tbl:BPTagnExamples}
}
\end{deluxetable*}

One common challenge to most fits is crowding. Figure~\ref{fig:agnExamples}, 
Panel~H shows a galaxy in the cluster A644 where many neighbors have been 
removed by subtraction and only the target was fit with GALFIT. Our algorithm 
determined that all neighbors were sufficiently far or faint enough that they 
did not interfere with the fitting procedure. To obtain a low value of 
\chisqnu, the pixels from these other objects in the field of view were simply 
masked out. Figure~\ref{fig:agnExamples}, Panel~D depicts a similar but 
slightly more challenging case. Here the target is blended with a bright, 
nearby object. In this case multiple objects are fit to obtain a reasonable 
model for the target. As before, objects yet further away are simply masked 
out. Panels~A~\&~B present additional examples of this case. A final case is 
illustrated by Figure~\ref{fig:BPTagnExamples}, Panel~M. In this case the 
model is simply inadequate because the galaxy's morphology is more complicated 
than our simple models. This galaxy appears to be both blended with other 
objects and morphologically disturbed. 

\begin{figure}
\epsscale{1.0}
\plotone{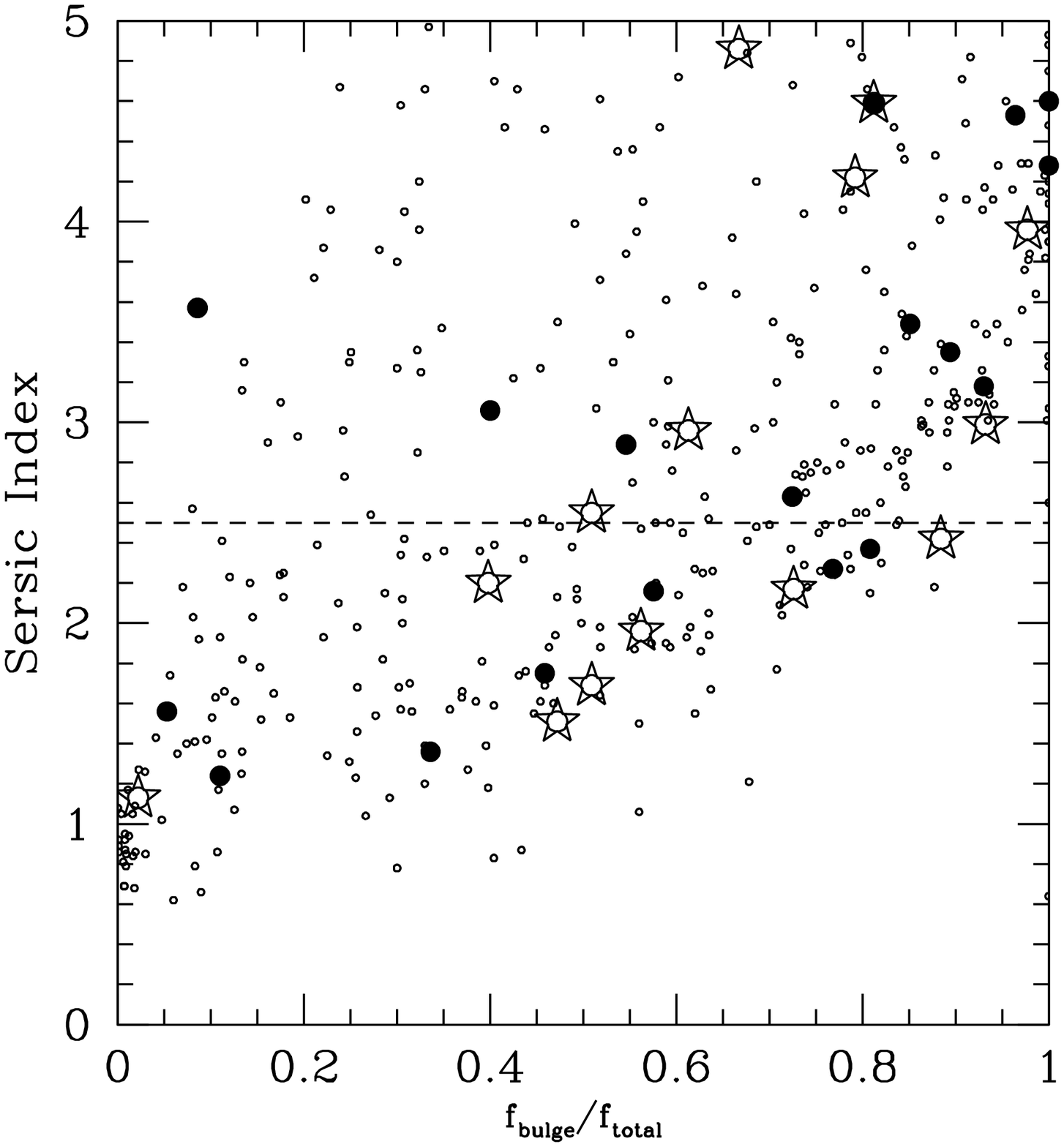}
\caption{Bulge-to-total flux ratio versus S\'ersic index for all of the groups 
and clusters in our sample. Points are as in Figure~\ref{fig:bpt} for 
X-ray AGN ({\it filled circles}), BPT AGN ({\it stars}), and inactive 
galaxies ({\it open circles}). This plot includes all the galaxies 
successfully fit with GALFIT, though the BPT AGN are only selected from the 
subsample with adequate Sloan spectra. The dashed line is drawn at $n=2.5$.
\label{fig:btvn}}
\end{figure}

The algorithm used by GALFIT is optimized for speed and is searching a very 
large parameter space; consequently the probability of getting lost in a local 
minimum is non-negligible. Though it is difficult to be certain that this has 
not happened for most cases, we employ several techniques to guard against 
this possibility. First, we run GALFIT on each target several times and update 
the initial fit parameters if the resultant \chisqnu\ differs by more than 
several percent between two runs. Second, we fit two distinct sets of models 
to each galaxy in separate runs: the S\'ersic index in a single component fit 
and the bulge-to-total flux ratio in a two component model fit. These 
quantities are correlated for most galaxies (see Figure~\ref{fig:btvn}) and we 
carefully reexamine egregious outliers. Finally, we save output images, as 
shown in Figures \ref{fig:agnExamples} and \ref{fig:BPTagnExamples}, and 
inspect these to identify poor fits. 

Stubborn objects that are not well fit by our procedure persist, though
they are relatively few. As noted above, GALFIT was originally developed by 
\citet{peng02} with the purpose of decomposing the complex structure of well 
resolved galaxies. We instead use it to do simple one- or, at most, two-model 
fits. For our typical resolution and galaxy size this is not a problem.
However, some of the galaxies in our images are so well resolved that a simple
S\'ersic profile or de~Vaucouleurs bulge plus exponential disk fit is 
insufficient to adequately describe the morphology. Structures such as bars 
or prominent spiral arms are sometimes fit rather than the more averaged 
profile of the galaxy that may have resulted from a less well-resolved image 
of the same object. Also, we occasionally find objects that are blended 
or coincident with the target and cannot be simultaneously with the target, 
such as the example of Panel~M of Figure~\ref{fig:BPTagnExamples} mentioned 
above. In general, blending is a particularly challenging problem if 
\textsc{SExtractor} fails to find the blended object as a distinct source.

\begin{figure}
\plotone{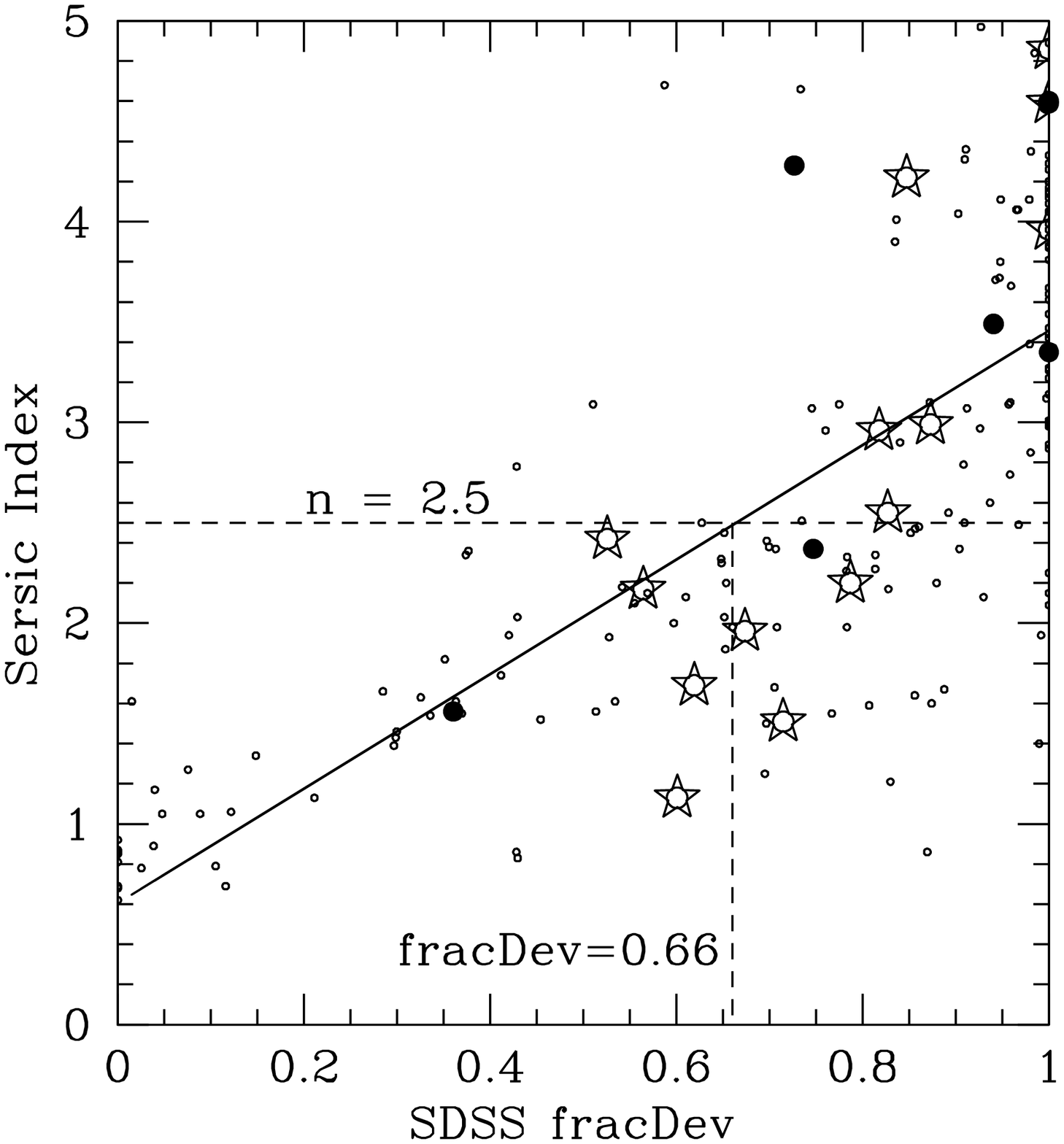}
\caption{S\'ersic index versus SDSS \texttt{fracDeV} parameter for all of the 
galaxies with SDSS coverage. Points are as in Figure~\ref{fig:bpt} for
X-ray AGN ({\it filled circles}), BPT AGN ({\it stars}), and inactive
galaxies ({\it open circles}). The solid line is a linear fit and the dashed
lines are drawn at $n=2.5$, where we separate early and late-type galaxies,
and the interception of the fit line with $n=2.5$, at \texttt{fracDeV} $=0.66$.
\label{fig:nvfracDeV}}
\end{figure}

In addition to comparison of the correlation between the S\'ersic index and 
the bulge-to-total flux ratio shown in Figure~\ref{fig:btvn}, we also 
compare to the SDSS \texttt{fracDeV} parameter when available. We find that 
these two parameters do not correlate well, while our S\'ersic index 
measurements do correlate with the \texttt{fracDeV} parameter, as shown in 
Figure~\ref{fig:nvfracDeV}, although the scatter is significant. One source 
of scatter is that the \texttt{fracDeV} parameter saturates while we still 
measure a substantial 
range of  S\'ersic index. We fit a line to this relation, although excluded 
points with \texttt{fracDeV} $= 1$, and found that \texttt{fracDeV} $= 0.66$
corresponds to $n=2.5$. This is reasonably consistent with the value of 
\texttt{fracDeV} $= 0.8$ used in the literature to identify early-type 
galaxies \citep[e.g.][]{bernardi05}. As for Figure ~\ref{fig:btvn}, 
we visually inspected all outliers on this figure to check the goodness of fit.

\begin{deluxetable*}{lrrrrrrrrrrrr}
\tablecolumns{13}
\tabletypesize{\scriptsize}
\tablecaption{Morphological Results}
\tablehead{
\colhead{Cluster / Group} &
\multicolumn{2}{c}{--- $N_{\mathrm{fit}}$ ---} &
\colhead{$f_{n>2.5}$} &
\multicolumn{3}{c}{------ $N_{\mathrm{total}}$ ------} &
\multicolumn{2}{c}{--- $N_{\mathrm{fit}^*}$ ---} &
\colhead{$f_{n>2.5}^*$} &
\multicolumn{3}{c}{------ $N_{\mathrm{total}}^*$ ------} \\
\colhead{} & 
\colhead{$All$} & 
\colhead{$n>2.5$} & 
\colhead{} &
\colhead{$Raw$} & 
\colhead{$Corr$} & 
\colhead{$n>2.5$} &
\colhead{$All$} & 
\colhead{$n>2.5$} & 
\colhead{} & 
\colhead{$Raw$} & 
\colhead{$Corr$} & 
\colhead{$n>2.5$} \\
\colhead{(1)} & \colhead{(2)} & \colhead{(3)} &
\colhead{(4)} & \colhead{(5)} & \colhead{(6)} &
\colhead{(7)} & \colhead{(8)} & \colhead{(9)} &
\colhead{(10)} & \colhead{(11)} & \colhead{(12)} &
\colhead{(13)}
}
\startdata
A85              & 104 &  76 & 0.731$^{+0.094}_{-0.084}$ & 109  & 109  &  80 &  49   &  41  &  0.837$^{+0.152}_{-0.130}$  &  53  &  53  &  44  \\ 
A644             &  15 &   7 & 0.467$^{+0.278}_{-0.231}$ &  19  &  75  &  35 &   6   &   5  &  0.833$^{+0.167}_{-0.360}$  &  40  &  40  &  33  \\
A3128            &  54 &  21 & 0.389$^{+0.105}_{-0.084}$ &  67  &  67  &  26 &  25   &  15  &  0.600$^{+0.198}_{-0.153}$  &  28  &  28  &  17  \\
RXCJ0110.0+1358  &  30 &  19 & 0.633$^{+0.181}_{-0.144}$ &  30  &  30  &  19 &  15   &  12  &  0.800$^{+0.200}_{-0.228}$  &  15  &  15  &  12  \\     
RXCJ0746.6+3100  &  23 &  17 & 0.739$^{+0.226}_{-0.177}$ &  23  &  23  &  17 &  16   &  14  &  0.875$^{+0.125}_{-0.231}$  &  16  &  16  &  14  \\          
RXCJ1022.0+3830  &  36 &  11 & 0.306$^{+0.123}_{-0.091}$ &  36  &  36  &  11 &  18   &   6  &  0.333$^{+0.199}_{-0.132}$  &  18  &  18  &   6  \\          
All Clusters     & 262 & 151 & 0.576$^{+0.051}_{-0.047}$ & 284  & 340  & 188 & 129   &  93  &  0.721$^{+0.083}_{-0.075}$  & 170  & 170  & 126  \\   
 & & & & & & & & & & & &  \\
A3125            &  18 & 10 & 0.556$^{+0.237}_{-0.173}$ &  20  &  28  &  16  & 11  &  10  &  0.909$^{+0.091}_{-0.283}$  &  15  &  15  & 14  \\
A89B             &  14 &  7 & 0.500$^{+0.269}_{-0.184}$ &  22  &  22  &  11  &  7  &   5  &  0.714$^{+0.286}_{-0.308}$  &  12  &  12  &  9  \\ 
RXCJ0844.9+4258  &  13 & 10 & 0.769$^{+0.231}_{-0.239}$ &  13  &  13  &  10  &  9  &   8  &  0.889$^{+0.111}_{-0.308}$  &   9  &   9  &  8  \\          
RXCJ1002.6+3241  &  33 & 16 & 0.485$^{+0.154}_{-0.120}$ &  33  &  33  &  16  &  9  &   6  &  0.667$^{+0.333}_{-0.264}$  &   9  &   9  &  6  \\          
RXCJ1122.2+6712  &  22 &  8 & 0.364$^{+0.179}_{-0.126}$ &  22  &  22  &   8  &  8  &   5  &  0.625$^{+0.375}_{-0.270}$  &   8  &   8  &  5  \\          
RXCJ1204.4+0154  &  12 &  9 & 0.750$^{+0.250}_{-0.245}$ &  12  &  12  &   9  &  7  &   6  &  0.857$^{+0.143}_{-0.340}$  &   7  &   7  &  6  \\          
RXCJ1223.1+1037  &   4 &  2 & 0.500$^{+0.500}_{-0.323}$ &   4  &   4  &   2  &  2  &   1  &  0.500$^{+0.500}_{-0.414}$  &   2  &   2  &  1  \\          
RXCJ1324.1+1358  &   6 &  2 & 0.333$^{+0.440}_{-0.215}$ &   6  &   6  &   2  &  3  &   2  &  0.667$^{+0.333}_{-0.431}$  &   3  &   3  &  2  \\          
RXCJ1440.6+0328  &  15 &  7 & 0.467$^{+0.251}_{-0.172}$ &  15  &  15  &   7  &  9  &   5  &  0.556$^{+0.376}_{-0.240}$  &   9  &   9  &  5  \\          
RXCJ1604.9+2355  &   9 &  7 & 0.778$^{+0.222}_{-0.287}$ &   9  &   9  &   7  &  3  &   3  &  1.000$^{+0.000}_{-0.544}$  &   3  &   3  &  3  \\                 
All Groups       & 146 & 78 & 0.534$^{+0.068}_{-0.060}$ & 156  & 164  &  88  & 68  &  51  &  0.750$^{+0.120}_{-0.105}$  &  77  &  77  & 59 
\enddata
\tablecomments{Morphological and demographic information for our two samples. 
Columns are:
(1) Cluster or group name; 
(2) Number of objects successfully fit by GALFIT with $M_R \leq -20$ mag; 
(3) The number objects from column~2 with $n>2.5$; 
(4) Fraction of objects from column~2 with $n>2.5$; 
(5) Number of confirmed objects with $M_R \leq -20$ mag; 
(6) Number of objects after a completeness correction, if any; 
(7) Number of objects in the sample with $n>2.5$, corrected for completeness, 
if applicable; 
(8-13) Same as columns~2--7, but with the brighter magnitude cut 
($M_R = M_R^*$ + 1).
The errorbars on the fractions are all single-sided, one-sigma confidence 
intervals \citep{gehrels86}. 
\label{tbl:morphology}
}
\end{deluxetable*}

We performed this morphological analysis on all of the confirmed member 
galaxies and additional groups and clusters from \citep{sivakoff08}. 
We used these data to identify all of the early-type galaxies in each 
group and cluster with $M_R \leq -20$ mag and $M_R \leq M_R^*+1$ and 
calculate the early-type galaxy fraction. These results are presented in 
Table~\ref{tbl:morphology}, which includes the total number of galaxies 
and the number successfully fit. Typically these numbers agree, except for 
rare instances when some galaxies were not fit successfully. 

\section{AGN Fractions} \label{sec:AGNFractions}

The next step of our analysis is to combine the AGN classifications from 
\S\ref{sec:classification} and the morphology fits from 
\S\ref{sec:morphology} to measure the AGN fraction as a function of 
environment and determine if there is any variation with host galaxy 
morphology. This analysis is described in the first subsection below. 
In addition, in \S\ref{sec:classification} above we demonstrated that 
the X-ray AGN and BPT AGN were nearly disjoint populations. In the following 
subsection we compare the morphologies of these two populations. Throughout 
we calculate AGN fractions for absolute magnitude limits of $M_R \leq -20$ mag 
and $M_R \leq M_R^* + 1$ and in all cases the AGN fraction is defined to be 
the number of AGN divided by the total number galaxies above a given 
absolute magnitude limit. All errorbars are derived from Poisson and
binomial statistics 
and are single-sided, $1\sigma$ confidence intervals \citep{gehrels86}. 

\subsection{X-ray AGN} \label{sec:xagn} 

\begin{figure}
\epsscale{1.0}
\plotone{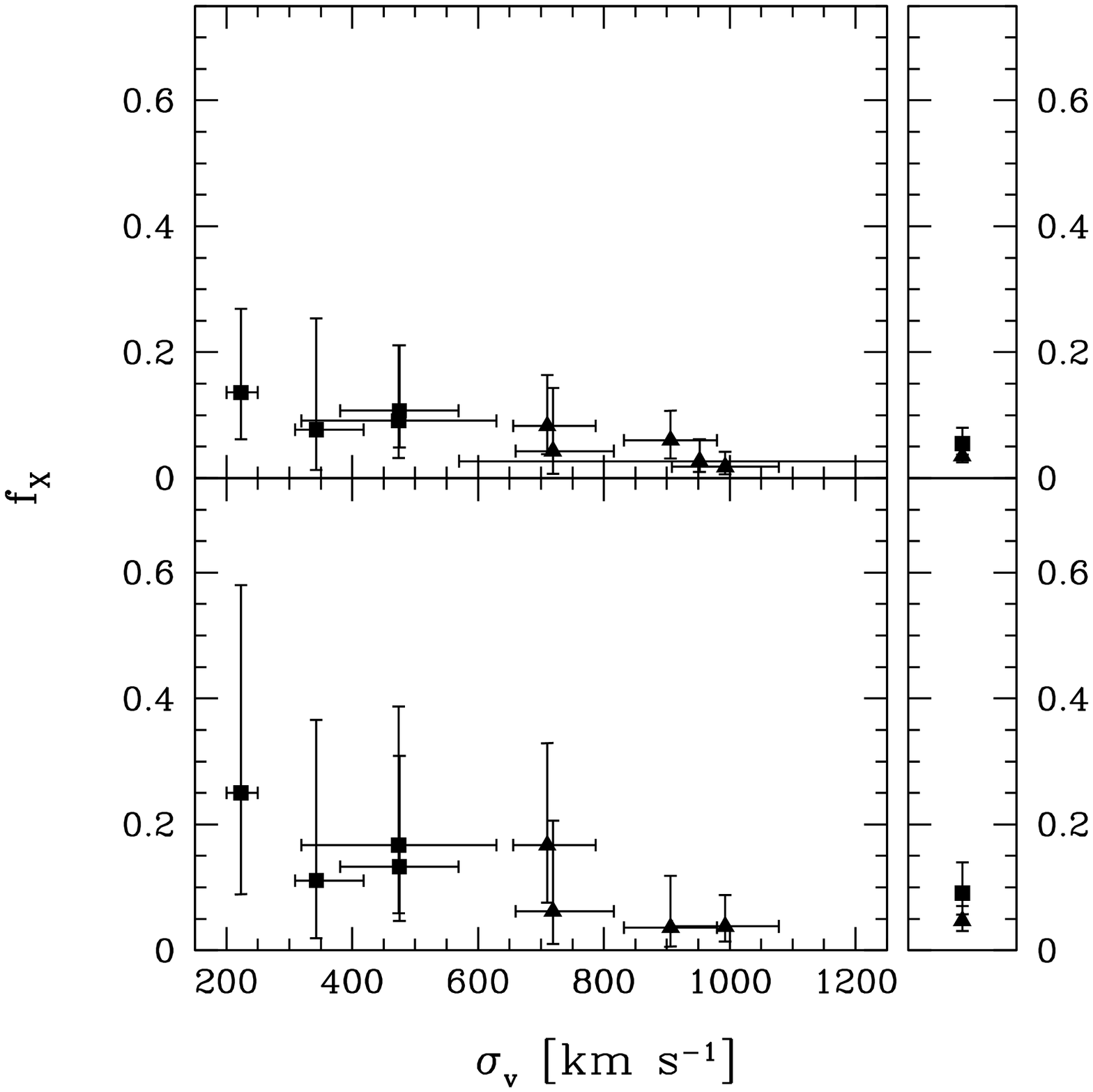}
\caption{X-ray AGN fractions versus velocity dispersion. The top panel of
this figure shows this relationship for an $M_R \leq -20$ cut on host galaxy 
absolute magnitude, as in \citet{martini06} and \citep{sivakoff08}. The bottom 
panel uses the cutoff of $M_R = M_R^* + 1$, as in \citet{martini09}. 
Measurements for all groups with $\sigma < 500$ \kms\ ({\it filled squares}) 
and clusters ({\it filled triangles}) with at least one AGN are shown. 
Following \citet{sivakoff08}, the right panels show the average AGN fraction 
for $\sigma < 500$ km~s$^{-1}$\ ({\it squares}), $\sigma > 500$ km~s$^{-1}$\ 
({\it triangles}). 
\label{fig:Xray_versus_sigma}}
\end{figure}

\begin{figure}
\epsscale{1.0}
\plotone{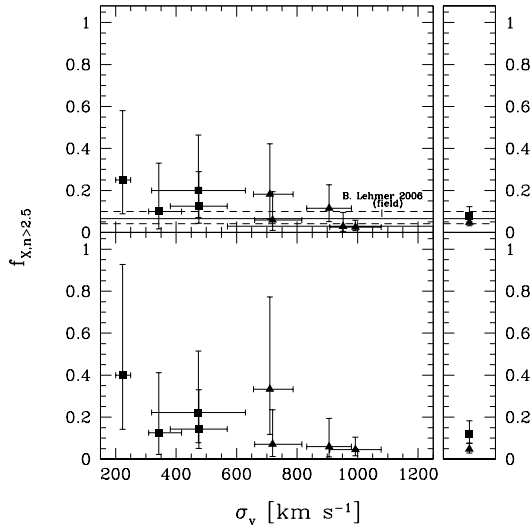}
\caption{As in Figure~\ref{fig:Xray_versus_sigma} for the early type X-ray AGN 
fractions versus velocity dispersion. The top panel of this figure shows a 
measurement of the field early-type X-ray AGN fraction from \citet{lehmer06}. 
\label{fig:earlytypeXray_v_sigma}}
\end{figure}

Table~\ref{tbl:xrayfrac} provides the number of X-ray identified AGN in each 
group or cluster, the X-ray AGN fraction, and the X-ray AGN fraction with 
early-type host galaxies. These results are also illustrated graphically 
in Figures~\ref{fig:Xray_versus_sigma} and \ref{fig:earlytypeXray_v_sigma}, 
although groups and clusters with no AGN are not shown for clarity. 
In both figures the top panel shows the results for the absolute magnitude 
threshold of $M_R \leq -20$ mag and the bottom panel for $M_R \leq M_R^*+1$. 
These figures indicate that the AGN fraction is smaller in environments 
characterized by a higher velocity dispersion. 

\begin{deluxetable*}{lrrrrrrrr}
\tablecolumns{9}
\tabletypesize{\scriptsize}
\tablecaption{X-ray AGN Fractions}
\tablehead{
\colhead{Cluster/Group} &
\colhead{$N_{X,fit,n>2.5}$} &
\colhead{$N_{X,fit}$} &
\colhead{$f_{X}$} &
\colhead{$f_{X,n>2.5}$} &
\colhead{$N_{X,fit,n>2.5}^*$} &
\colhead{$N_{X,fit}^*$} &
\colhead{$f_{X}^*$} &
\colhead{$f_{X,n>2.5}^*$} \\
\colhead{(1)} &
\colhead{(2)} &
\colhead{(3)} &
\colhead{(4)} &
\colhead{(5)} &
\colhead{(6)} &
\colhead{(7)} &
\colhead{(8)} &
\colhead{(9)}
}
\startdata
A85               &  2  &   2  & 0.018$^{+0.024}_{-0.012}$  & 0.025$^{+0.033}_{-0.016}$ &   2   &   2   &   0.038$^{+0.050}_{-0.024}$  &  0.045$^{+0.060}_{-0.029}$  \\
A644              &  1  &   2  & 0.027$^{+0.035}_{-0.017}$  & 0.029$^{+0.066}_{-0.024}$ &   0   &   1   &   0.025$^{+0.057}_{-0.021}$  &  0.000$^{+0.056}_{-0.000}$  \\
A3128             &  3  &   4  & 0.060$^{+0.047}_{-0.029}$  & 0.115$^{+0.112}_{-0.063}$ &   1   &   1   &   0.036$^{+0.082}_{-0.030}$  &  0.059$^{+0.135}_{-0.049}$  \\
RXCJ0110.0+1358   &  0  &   0  & 0.000$^{+0.061}_{-0.000}$  & 0.000$^{+0.097}_{-0.000}$ &   0   &   0   &   0.000$^{+0.123}_{-0.000}$  &  0.000$^{+0.153}_{-0.000}$  \\    
RXCJ0746.6+3100   &  1  &   1  & 0.043$^{+0.100}_{-0.036}$  & 0.059$^{+0.135}_{-0.049}$ &   1   &   1   &   0.062$^{+0.144}_{-0.052}$  &  0.071$^{+0.164}_{-0.059}$  \\    
RXCJ1022.0+3830   &  2  &   3  & 0.083$^{+0.081}_{-0.045}$  & 0.182$^{+0.240}_{-0.117}$ &   2   &   3   &   0.167$^{+0.162}_{-0.091}$  &  0.333$^{+0.440}_{-0.215}$  \\    
All Clusters      &  9  &  12  & 0.035$^{+0.013}_{-0.010}$  & 0.048$^{+0.022}_{-0.016}$ &   6   &   8   &   0.047$^{+0.023}_{-0.016}$  &  0.048$^{+0.028}_{-0.019}$  \\
& & & & & & & & \\
A3125             &  2  &   3  & 0.107$^{+0.104}_{-0.058}$  & 0.125$^{+0.165}_{-0.081}$ &   2   &   2   &   0.133$^{+0.176}_{-0.086}$  &  0.143$^{+0.188}_{-0.092}$  \\
A89B              &  2  &   2  & 0.091$^{+0.120}_{-0.059}$  & 0.182$^{+0.240}_{-0.117}$ &   2   &   2   &   0.167$^{+0.220}_{-0.108}$  &  0.222$^{+0.293}_{-0.144}$  \\
RXCJ0844.9+4258   &  1  &   1  & 0.077$^{+0.177}_{-0.064}$  & 0.100$^{+0.230}_{-0.083}$ &   1   &   1   &   0.111$^{+0.255}_{-0.092}$  &  0.125$^{+0.287}_{-0.103}$  \\    
RXCJ1002.6+3241   &  0  &   0  & 0.000$^{+0.056}_{-0.000}$  & 0.000$^{+0.115}_{-0.000}$ &   0   &   0   &   0.000$^{+0.205}_{-0.000}$  &  0.000$^{+0.307}_{-0.000}$  \\    
RXCJ1122.2+6712   &  2  &   3  & 0.136$^{+0.133}_{-0.074}$  & 0.250$^{+0.330}_{-0.161}$ &   2   &   2   &   0.250$^{+0.330}_{-0.161}$  &  0.400$^{+0.527}_{-0.258}$  \\    
RXCJ1204.4+0154   &  0  &   0  & 0.000$^{+0.153}_{-0.000}$  & 0.000$^{+0.205}_{-0.000}$ &   0   &   0   &   0.000$^{+0.263}_{-0.000}$  &  0.000$^{+0.307}_{-0.000}$  \\    
RXCJ1223.1+1037   &  0  &   0  & 0.000$^{+0.460}_{-0.000}$  & 0.000$^{+0.920}_{-0.000}$ &   0   &   0   &   0.000$^{+0.920}_{-0.000}$  &  0.000$^{+1.000}_{-0.000}$  \\    
RXCJ1324.1+1358   &  0  &   0  & 0.000$^{+0.307}_{-0.000}$  & 0.000$^{+0.920}_{-0.000}$ &   0   &   0   &   0.000$^{+0.614}_{-0.000}$  &  0.000$^{+0.920}_{-0.000}$  \\        
RXCJ1440.6+0328   &  0  &   0  & 0.000$^{+0.123}_{-0.000}$  & 0.000$^{+0.263}_{-0.000}$ &   0   &   0   &   0.000$^{+0.205}_{-0.000}$  &  0.000$^{+0.368}_{-0.000}$  \\        
RXCJ1604.9+2355   &  0  &   0  & 0.000$^{+0.205}_{-0.000}$  & 0.000$^{+0.263}_{-0.000}$ &   0   &   0   &   0.000$^{+0.614}_{-0.000}$  &  0.000$^{+0.614}_{-0.000}$  \\        
All Groups        &  7  &   9  & 0.055$^{+0.025}_{-0.018}$  & 0.080$^{+0.043}_{-0.029}$ &   7   &   7   &   0.091$^{+0.049}_{-0.034}$  &  0.119$^{+0.064}_{-0.044}$
\enddata
\tablecomments{X-ray AGN fractions for the cluster and group samples. 
Columns are: 
(1) Cluster or group name; 
(2) Number of X-ray identified AGN with good fits, an early-type morphology 
($n>2.5$), and $M_R \leq -20$ mag; 
(3) Number of all X-ray identified AGN with good fits in the sample; 
(4) Fraction of the fit galaxies that are X-ray identified AGN; 
(5) Fraction of the fit galaxies that are X-ray AGN and have early-type 
morphologies ($n>2.5$); 
(6--9) Same as (2--5) but with the brighter absolute magnitude cut of 
$M_R \leq M_R^* + 1$. 
\label{tbl:xrayfrac}
}
\end{deluxetable*}

Due to the small number of AGN in individual groups and clusters, the 
statistical significance of these trends is difficult to discern. We thus 
bin the data for all groups and all clusters separately, where we divide the 
two samples at a velocity dispersion of $\sigma = 500$~\kms, and compare these 
two environments. The binned results are presented in Table~\ref{tbl:xrayfrac} 
and the right-hand panels of Figures~\ref{fig:Xray_versus_sigma} and 
\ref{fig:earlytypeXray_v_sigma}. These results make the trend with velocity 
dispersion more clear. For the higher-luminosity host galaxies 
($M_R < M_R^*+1$), the errorbars on the binned AGN fractions for the groups 
and clusters do not overlap for both all AGN and just those with early-type 
hosts. Specifically, for all galaxies more luminous than $M_R^*+1$ we find 
that the X-ray AGN fraction is $f_A = 0.091^{+0.049}_{-0.034}$ for groups and 
$0.047^{+0.023}_{-0.016}$ for clusters, or a factor of two 
higher in groups. The trend is somewhat more pronounced for the early-type 
galaxies, where the AGN fraction is $f_{A,n>2.5} = 0.119^{+0.064}_{-0.044}$ in 
groups and $0.048^{+0.028}_{-0.019}$ for clusters. This demonstrates that the 
AGN fraction is a factor of two higher in groups relative to clusters and that 
the AGN fraction is similarly higher when just early-type host galaxies are 
considered. 

The errorbars on these binned results, shown in the righthand panels of 
Figures~\ref{fig:Xray_versus_sigma}~and~\ref{fig:earlytypeXray_v_sigma}, show
the same single-sided, $1\sigma$ confidence intervals as the results for 
individual groups and clusters. In principle, if these binned error bars 
exactly overlap, then each population is distinct from the other with 
84\% confidence, and if not, then the confidence level can be obtained by 
expanding or contracting the confidence limits until they exactly overlap. 
However, there is additional uncertainty due to the choice of
$\sigma = 500$~\kms\ to separate groups and clusters. While this choice was physically 
motivated and is not an unreasonable point to divide the sample, there are 
additional, physically meaningful values of the velocity dispersion to 
separate groups and clusters. For example, we could have chosen to divide 
groups and clusters at $\sigma = 400$~\kms\ instead of 500~\kms. Therefore,
proper statistical analysis of the difference between the two samples needs 
to include an additional penalty that reflects the other options 
for binning the data. We account for this by raising the previous probability 
to an exponent that represents all of the other ways we could have binned the 
data. Thus, the probability that two populations are distinct from one another 
is
 \begin{equation}
     (1-(1-\mathrm{CL})^2))^{N_\mathrm{groupings}},
 \end{equation}
where CL is the confidence limit at which the error bars just overlap, and
$N_\mathrm{groupings}$ is the number of possible groupings. For example, a
confidence limit of 84\% (one-sigma) and three possible groupings 
yields a 93\% probability that the two populations are distinct. Note that in 
\citet{sivakoff08} a similar analysis was performed to show that groups and 
clusters were different from one another, but there all possible combinations 
of the data were considered. In our case there are in principle ${16 \choose 2} 
= 120$ possible ways to create two samples out of our data, 
but we do not adopt this approach as most of these options are nonconsecutive 
and not physically motivated. From this analysis we conclude that the X-ray AGN 
and early-type X-ray AGN fractions are higher in groups relative to clusters 
at the 85\% and 92\% level, respectively, for the $M_R \leq M_R^*+1$ sample. 
The statistical significance for the $M_R \leq -20$ mag sample is lower 
(77\% and 76\%, respectively), although the trend is consistent. 

Another intriguing question we can begin to answer is how the AGN fractions 
in groups and clusters compare to the field value. While there is little 
data on X-ray selected AGN fractions in the field, the study of 
\citet{lehmer06} examined the X-ray fraction in early-type galaxies as a 
function of redshift. When calculated with the same selection criteria 
as we employ, the field early-type AGN fraction is 6.6\% for $M_R \leq -20$ mag 
\citep[B. Lehmer, private communication, see also][]{martini07}. 
Intriguingly, the group and cluster early-type AGN fractions are both 
consistent within the errorbars with the field value, although the cluster 
fraction is lower and the group fraction is higher. 

\subsection{Emission-line and X-ray AGN} \label{sec:bptagn} 

The eight groups and three poor clusters 
selected from the NORAS sample have SDSS 
spectra as well as imaging data. Consequently, we were also able to classify 
them as BPT AGN or not based on their visible-wavelength emission line ratios 
(see \S~\ref{sec:bptClassification}). We calculate the BPT 
AGN fraction as the ratio of BPT AGN in galaxies more luminous than 
$M_R = -20$ mag divided by the total number of galaxies above this luminosity 
with SDSS spectra. As noted above, there are 14 AGN above the \citet{kewley01} 
classification line of 199 total galaxies with spectra. Six of these BPT AGN 
are in 88 cluster galaxies and eight are in 111 group galaxies. The 
corresponding BPT AGN fractions are $0.068^{+0.041}_{-0.027}$ for groups and 
$0.072^{+0.036}_{-0.025}$ for clusters. These fractions are consistent. 
Only one of these 14 BPT AGN is also classified as an X-ray AGN and it is a 
member of a group (RXCJ0844.9+4258). For comparison the X-ray AGN fraction is 
$0.039^{+0.019}_{-0.014}$, although is drawn from a larger host galaxy 
population than the two BPT fractions quoted above. Our BPT AGN fraction is 
comparable but somewhat lower than the X-ray AGN fraction of $f_A \sim 0.07$ 
(based on one object) found by \citet{shen07} for $M_R \leq -20$ mag. Their 
BPT AGN fraction is also $\sim7$\%, although this was calculated for 
$M_R \leq -18$ mag. At this lower threshold they measure an X-ray AGN fraction 
of $0.7$\% (one out of $\sim 140$ galaxies). 
We similarly find BPT AGN that are not classified as 
X-ray AGN and X-ray AGN that are not classified as BPT AGN. 

\begin{figure}
\epsscale{1.0}
\plotone{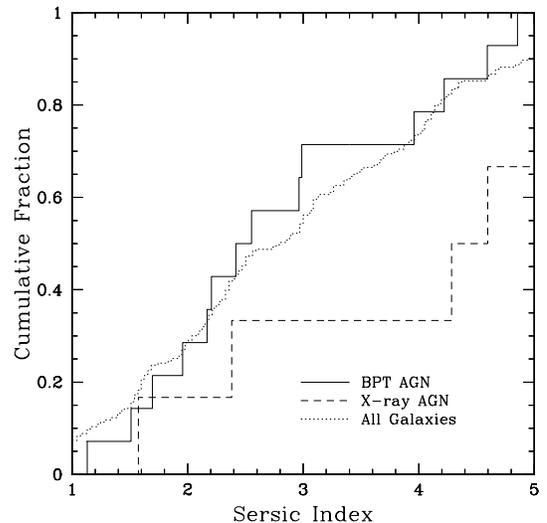}
\caption{Cumulative fraction of galaxies with S\'ersic index $n$ less than the 
amount shown for the sample selected from the NORAS catalog. The BPT AGN 
sample ({\it solid histogram}) includes a larger fraction with small S\'ersic 
index, corresponding to later-type morphologies, while the X-ray AGN ({\it 
dashed histogram}) hosts are predominantly early-type galaxies; however, the 
difference is not statistically significant. All other galaxies with 
spectra and morphological fits are also shown ({\it dotted histogram}). There 
are 14 BPT AGN and six X-ray AGN in these histograms. The last two X-ray AGN 
are at $n = 5$ and one AGN is common to both distributions. 
\label{fig:BPTvXrayMorphologies}}
\end{figure}

We also compare the morphologies of the BPT AGN and the X-ray AGN to determine
if the lower fraction of high-luminosity AGN in denser regions found in 
SDSS \citep{kauffmann04} is correlated with the morphology--density relation. 
To make this comparison we plot the cumulative fraction of X-ray and BPT 
AGN as a function of S\'ersic index in Figure~\ref{fig:BPTvXrayMorphologies}. 
While this figure may suggest that the X-ray AGN are more likely to be in 
early-type host galaxies, both a KS test and a Mann-Whitney $U-$Test indicate
that the samples are consistent. The BPT AGN morphologies are also in very 
good agreement with the galaxies not classified as AGN by either method. The 
X-ray and BPT AGN shown in this figure are only the subset with spectroscopy 
from SDSS, and therefore not all of the X-ray AGN are shown. 

\section{Discussion} \label{sec:discussion}
The results of the previous sections have shown that there is
tentative evidence that the AGN fraction is 
lower in clusters than in groups, and that this difference also holds when 
only early-type galaxies are considered separately. This second statement 
is important because it helps to break the degeneracy between morphology and 
density and between morphology and AGN fraction. Thus the larger fraction of 
AGN in groups indicates that the AGN fraction is higher in both early 
and late-type galaxies and is not simply due to a larger fraction of late-type 
galaxies in lower velocity dispersion environments. These results have 
interesting implications for AGN triggering and fueling, as well as the 
evolution of galaxies in groups and clusters. If true, one of the main 
implications is 
that early-type galaxies in groups are more likely to be AGN than their 
counterparts in clusters. This could be explained by higher cold gas fractions 
or greater likelihood of triggering, such as due to an interaction; most likely 
both play a role. In clusters, the higher rates of galaxy interactions and gas 
stripping could remove potential fuel from the galaxies more efficiently than 
those same interactions driving angular momentum loss and AGN fueling. It has
also been proposed \citep{hopkins08a,hopkins08b} that smaller, less massive
groups are the ideal environment for AGN activity caused by mergers. Though these
studies focus on higher-luminosity AGN and rely on a different triggering
mechanism than in this work, it would be interesting to investigate whether 
our weakly observed trend 
continues to even lower-density group environments discussed by \citet{hopkins08a}.

These results also help to explain the range of conclusions that have been 
drawn about the AGN fraction as a function of environment. For example, 
\citet{dressler85} found the BPT AGN fraction in clusters is a factor of 
five times lower than in the field. Many studies have confirmed this
result for higher-luminosity AGN, yet found less of a difference at lower 
luminosities \citep{shimada00,miller03,kauffmann04,grogin05}. In contrast, 
X-ray observations of clusters find many more AGN than via the BPT technique 
\citep{martini06,martini07}, while observations of optically-selected, 
poor groups of galaxies identify a larger fraction of BPT AGN relative 
to X-ray AGN \citep{shen07}. 
Our spectroscopic analysis of the rich groups and poor clusters selected from 
the NORAS catalog demonstrates that much of these pronounced differences in 
the literature are due to differences in selection because the BPT AGN and 
X-ray AGN are nearly disjoint populations. There are substantial populations 
of both types of AGN in groups, while this is less often the case in higher 
galaxy density environments. The very low density groups in the \citet{shen07} 
sample are unlikely to be virialized systems, even though they are typical 
of groups found via redshift surveys such as SDSS. The X-ray emission from 
the more massive groups and poor clusters suggests that these are virialized 
systems, and the change between mostly BPT AGN and X-ray AGN may reflect 
a change in the dominant accretion mode between unvirialized and virialized 
systems. This is further supported by the cluster sample of \citet{martini06}, 
who found few BPT AGN and none that were not also detected as X-ray AGN, 
although many of these spectra were also of low signal-to-noise. For 
comparison, we find approximately equal numbers of BPT AGN in groups and 
clusters, including examples of BPT AGN in clusters that are not X-ray AGN. 
Nevertheless, from the several previous studies \citep{dressler99,martini06} 
the trend is that while the AGN population is lower in clusters than in groups, 
the decrease is larger for BPT AGN than for X-ray AGN. The more constant X-ray 
AGN fraction with local density is also similar to the trend seen in radio AGN 
by \citet{best05}, who found that the fraction of radio AGN is relatively 
insensitive to environment, compared to BPT AGN. 

A potential physical explanation of these trends is that the BPT and X-ray 
AGN trace different accretion modes. The BPT AGN exhibit line ratios 
characteristic of AGN whose spectral energy distributions are well-matched 
by thin disk models where the accretion rate is greater than $\sim 1$\% of the 
Eddington rate. In contrast, the weak or absent emission lines, at least 
in these moderate signal-to-noise spectra, combined with their substantial 
X-ray luminosities, suggest radiatively-inefficient accretion with lower 
accretion rates relative to Eddington \citep{narayan98,ho99,vasudevanfabian07}. 
This simple picture is supported by 
the weak trend that the X-ray AGN are more often found in early-type hosts than 
the BPT AGN. As both populations have comparable total luminosities, the 
X-ray AGN hosts have larger spheroids and are expected to have larger 
black hole masses as a result. As radio and X-ray emission are reasonably 
well correlated in these radiatively inefficient models 
\citep[e.g.][]{merloni03}, this hypothesis is also consistent with the 
distribution of the observed radio AGN fraction. 

Another interesting implication of this work is that the average AGN fraction 
for the groups and clusters together is similar to the field early-type 
X-ray AGN fraction of $6.6^{+3.4}_{-2.4}$\% (B.~Lehmer 2006, private 
communication), although the groups alone have a higher fraction. One 
interpretation of this result is that while galaxies in the field typically 
have substantial supplies of cold gas, perhaps even more than found in group 
galaxies, this is offset by a lower rate of triggering due to the lower 
density, at least if interactions and mergers play an important role. 
This implies that the cluster environment is too dense, the field environment 
is too sparse, and the group environment is ``just right'' to fuel and trigger 
X-ray AGN. Further observations are required to draw firm conclusions about the 
relative AGN fraction in the field, groups, and clusters, yet if the variation 
we find is confirmed it would strongly point to the importance of galaxy 
interactions (although not necessarily mergers) for fueling even 
lower-luminosity AGN. This would be surprising, as it conflicts with 
current studies based on companion fractions \citep{fuentes88,schmitt01,
li08}, although those studies are all based on BPT AGN, rather than X-ray 
AGN.


\section{Conclusions} \label{sec:summary}

The distribution of AGN as a function of environment is a potentially 
valuable probe of the fueling and triggering of AGN, as well as the 
connection between galaxy and black hole evolution. We have conducted a 
new survey of eleven rich groups and poor clusters selected from the 
NORAS catalog at $0.02 < z < 0.06$ to measure the AGN fraction as a function 
of environment. Our group sample, defined to be environments with $\sigma < 
500$~\kms, contains eight new groups plus two previously published in 
\citet{sivakoff08} and thus represents a factor of five increase in sample 
size. The cluster sample contains three new clusters and three previously 
published, or a factor of two increase in sample size. 

We identify X-ray AGN in these groups and clusters with a combination of 
spectral fits and flux ratio arguments to demonstrate that the X-ray emission 
from each X-ray AGN is inconsistent with other sources of X-ray emission. The 
main result of this analysis is that the X-ray AGN fraction is approximately 
a factor of two higher in groups than in clusters. This trend is apparent for 
both AGN in host galaxies more luminous than $M_R = -20$ mag and for more 
luminous hosts with $M_R \leq M_R^*+1$, although the difference has greater 
statistical significance for the higher luminosity threshold. The X-ray AGN 
fractions for the higher threshold are $f_A(L_X \geq 10^{41}; 
M_R \leq M_R^* + 1) = 0.047^{+0.023}_{-0.016}$ for clusters and 
$0.091^{+0.049}_{-0.034}$ for 
groups. There is a 85\% probability that the group AGN fraction is larger than 
the cluster AGN fraction. This result may be more significant for the more 
luminous galaxy subsample due to the larger fraction of the most luminous 
galaxies that are X-ray AGN. 

Because the incidence of AGN in galaxies depends on host galaxy morphology, 
and the distribution of galaxy morphologies depends on environment, we 
have conducted the first quantitative morphological analysis of the AGN 
fraction in dense environments with these six clusters and ten groups. 
The morphological data for every confirmed group and cluster galaxy was 
obtained with the GALFIT software package by \citet{peng02} and used to 
separate early-type and late-type galaxies. We then calculated the X-ray
AGN fraction for just the early-type galaxy populations in the groups and 
clusters separately and found that the early-type AGN fraction is also a 
factor of two higher in groups relative to clusters. For the higher galaxy 
luminosity threshold we find $f_{A,n>2.5}(L_X \geq 10^{41}; M_R \leq M_R^*+1) = 
0.048^{+0.028}_{-0.019}$ for clusters and $0.119^{+0.064}_{-0.044}$ 
for groups (92\% confidence).
The similar trends for early-type galaxies and all galaxies indicate that 
the AGN fraction is not different simply because the morphological mix of 
galaxies changes as a function of environment, but rather that all galaxy 
types have a higher probability of hosting an AGN in the group environment. 
In addition, the group value is also higher than the best estimate of the 
early-type AGN fraction in the field. 
This may be because group galaxies, even early-type group 
galaxies, are more likely to have substantial cold gas reservoirs for AGN 
fueling than cluster galaxies, while galaxy interactions are more likely to 
occur in groups than the field, or some combination of these effects. 

Finally, we have also estimated the AGN fraction based on emission-line 
diagnostics for the subset of the galaxies with SDSS spectroscopy. There are 
14 BPT AGN in this subset of groups and clusters, as well as six of our nine 
X-ray AGN. Strikingly, these populations are nearly 
completely disjoint: only one AGN meets our criterion as both an X-ray AGN and 
as a BPT AGN. This is a clear demonstration of how different selection 
techniques may identify different populations of AGN. Our morphological 
analysis shows that the host galaxies of these two AGN types are marginally 
different in that the X-ray AGN are more likely to be hosted by early-type 
galaxies. While the host galaxies for both AGN populations are more luminous 
than $M_R \leq -20$ mag, the earlier-type hosts for the X-ray AGN imply 
relatively larger supermassive black holes compared to the BPT AGN. These 
observations are thus consistent with lower-efficiency, but relatively 
more X-ray bright, accretion in the X-ray AGN. 

\acknowledgements 

We thank Greg Sivakoff for many helpful discussions. JSM acknowledges partial 
support for this work from NASA grant NNX07AQ60G. Based on observations 
obtained with XMM-Newton, an ESA science mission with instruments and 
contributions directly funded by ESA Member States and NASA.

\end{document}